\documentclass[12pt,preprint]{aastex}

\newcommand{\logg} {\log g}
\newcommand{\Te} {T_{\rm eff}}

\newcommand{\msun} {$M_\odot$}

\newcommand\gta{\lower 0.5ex\hbox{$\buildrel > \over \sim\ $}} 
\newcommand\lta{\lower 0.5ex\hbox{$\buildrel < \over \sim\ $}} 

\shortauthors{Genest-Beaulieu & Bergeron}
\shorttitle{Atmospheric Parameters of SDSS White Dwarfs}

\begin{document}

\title{Comparison of Atmospheric Parameters Determined from
  Spectroscopy and Photometry for DA White Dwarfs in the Sloan Digital
  Sky Survey}

\author{C.~Genest-Beaulieu, \& P. Bergeron}

\affil{D\'epartement de Physique, Universit\'e de Montr\'eal,
  C.P.~6128, Succ.~Centre-Ville, Montr\'eal, QC H3C 3J7, Canada;
  genest@astro.umontreal.ca, bergeron@astro.umontreal.ca}

\begin{abstract}

We present a comparative analysis of atmospheric parameters obtained
with the so-called photometric and spectroscopic
techniques. Photometric and spectroscopic data for 1360 DA white
dwarfs from the Sloan Digital Sky Survey (SDSS) are used, as well as
spectroscopic data from the Villanova White Dwarf Catalog. We first
test the calibration of the $ugriz$ photometric system by using model
atmosphere fits to observed data. Our photometric analysis indicates
that the $ugriz$ photometry appears well calibrated when the SDSS to
AB$_{95}$ zeropoint corrections are applied. The spectroscopic
analysis of the same data set reveals that the so-called high-$\logg$
problem can be solved by applying published correction functions that
take into account 3D hydrodynamical effects. However, a comparison
between the SDSS and the White Dwarf Catalog spectra also suggests
that the SDSS spectra still suffer from a small calibration
problem. We then compare the atmospheric parameters obtained from both
fitting techniques and show that the photometric temperatures are
systematically lower than those obtained from spectroscopic data. This
systematic offset may be linked to the hydrogen line profiles used in
the model atmospheres. We finally present the results of an analysis
aimed at measuring surface gravities using photometric data only.

\end{abstract}

\keywords{stars: fundamental parameters --- techniques: photometric
  --- techniques: spectroscopic --- white dwarfs}

\section{Introduction}

Various methods have been developed over the years to measure the
atmospheric parameters --- effective temperature ($\Te$) and surface
gravity ($\logg$) --- of hydrogen-line DA white dwarfs (see
\citealt{BSL92} for a review). By far, the most commonly employed
method nowadays is the so-called spectroscopic technique, where the
profiles of the hydrogen Balmer lines are compared with the
predictions of detailed model atmospheres (\citealt{BSL92},
\citealt{vennes1997}, \citealt{LBH05}, \citealt{koesterVoss2009},
\citealt{Tremblay2011}, \citealt{Gianninas2011}, to name a few).
A similar approach can of course be applied to any type of white
dwarfs, such as DB stars \citep{voss2007,bergeron2011}. When the
spectroscopic lines vanish at low effective temperatures, however, one
must rely on the spectral energy distribution obtained from broad band
energy distributions, and apply the photometric technique where
measured magnitudes in various passbands are converted into average
fluxes and compared to predicted fluxes from model atmospheres. Such a
technique has been successfully applied in the context of cool white
dwarfs by \citet{BRL97,BLR01}, for example, using optical $BVRI$ and
infrared $JHK$ photometry, or by \citet{kilic2010} using $ugriz$ photometry in the
optical instead. In such cases, the effective temperature can be
measured directly, but the surface gravity can only be determined if
the distance to the star is known through parallax measurements for
instance \citep{BLR01}.  Both the photometric and spectroscopic techniques are
powerful methods since they can be applied routinely to large samples
of white dwarfs.

Because of the sensitivity of the Balmer lines to variations in both
$\Te$ and $\logg$, the {\it precision}\footnote{For completeness, the
  precision of a measurement refers to its reproducibility or
  repeatability, while the accuracy is related to the closeness of the
  measured parameter to its actual true value.} of the spectroscopic
technique is extremely high, in particular in the context of DA
stars. Indeed, \citet{LBH05} have determined, using multiple spectroscopic
measurements of the same stars, that the precision could be as high as
1.2\% in $\Te$ and 0.038 dex in $\logg$. While the spectroscopic
technique is arguably the most precise method for measuring
atmospheric parameters of white dwarf stars, it also depends heavily
on the details of the physics of line broadening, and the calculations
of atomic level populations. For instance, the improved Stark
profiles of \citet{TB09}, which include nonideal effects directly into
the line profile calculations, can yield differences as high as 1000 K
in $\Te$ and 0.1 dex in $\logg$, which are even larger than
the quoted precision of the spectroscopic technique! Even more
dramatic are the effects of 3D hydrodynamical calculations on the
predicted line profiles, compared to standard
model atmospheres calculated within the mixing-length theory (see
\citealt{TremblayIV} and references therein). So even though
the spectroscopic technique has a high degree of precision, it
may still lack a similar level of accuracy.

While the photometric technique is admittedly less precise than the
spectroscopic method, it has the definite advantage of being less
sensitive to the details of model atmosphere calculations since it
relies mostly on the emergent continuum fluxes. On the other hand, the
photometric technique depends heavily on the flux calibration 
used to convert observed magnitudes into observed average fluxes, as
discussed at length by \citet{HB06}. Ideally, one would like to
compare both the spectroscopic and photometric techniques for a large
ensemble of spectroscopic and photometric data for the same objects.

The large number of white dwarf stars identified in the Sloan Digital
Sky Survey (SDSS) offers such a unique opportunity to compare
atmospheric parameters derived from photometry, using the homogeneous
set of $ugriz$ photometric data, and from spectroscopy, using the
homogeneous set of medium resolution spectroscopic data, which are available {\it
  for the same objects}. Such a comparison has been performed by 
\citet[][see their Figure 16]{TremblayIV} but only for a limited
range of effective temperatures. Note also that \citet{eisenstein06} used both
photometric and spectroscopic data simultaneously to measure the
atmospheric parameters of the white dwarfs in the Data Release
4. Because we are interested here in comparing the results from both
techniques, and in particular the temperature scales, we will consider
these two data sets independently.

We thus present in this paper a detailed comparison of the photometric
and spectroscopic techniques applied to the DA stars identified in the
SDSS. The photometric sample used in our sample is described
and analyzed in Section 2, where we also explore the effects of
interstellar reddening and other issues related to flux calibration.
The optical spectra for the same objects are then analyzed in Section
3, together with independent white dwarf spectra drawn from the
Villanova White Dwarf Catalog, which are used to test the flux
calibration of the SDSS spectroscopic data. The effective temperatures
derived from both photometric and spectroscopic techniques are then
compared in Section 4, while in Section 5, we exploit the sensitivity
of the $u-g$ color index to $\logg$ to compare surface gravities
measured photometrically and spectroscopically.  Our summary and
conclusions follow in Section 6.

\section{Photometric Analysis}

\subsection{Photometric Sample}\label{sect:phot_sample}

Our photometric sample is based on the Data Release 7 (DR7) of the
Sloan Digital Sky Survey \citep{DR7}, which contains 19,713 white
dwarfs of various spectral types, including 12,831 DA stars. We
exclude from this sample all subtypes DAB, DAO, etc. Since the main
purpose of our analysis is to compare atmospheric parameters derived
from photometry and spectroscopy, we retain only the best data sets
for each technique, and we thus restrict our photometric sample to white
dwarfs with a signal-to-noise ratio above 25 in the $g$-band, and with
uncertainties less than 0.1 mag in all bandpasses. Applying
these criteria, we end up with a sample of 1478 $ugriz$ photometric
data sets for 1360 DA white dwarfs, given that some objects have
been observed more than once. Those are treated as independent
observations.

Our photometric sample is summarized in the $(u-g,\ g-r)$ two-color
diagram displayed in Figure \ref{fig:umg_gmr}. Also superimposed on
the observations are the theoretical photometric sequences for
hydrogen atmosphere white dwarfs from \citet{HB06}\footnote{See also
  http://www.astro.umontreal.ca/\~{ }bergeron/CoolingModels.}. Note
that the magnitudes displayed here have been corrected for
interstellar reddening following a procedure described in Section
\ref{sect:reddening}. One can already notice the sensitivity of the
$u-g$ color index to surface gravity between $\Te \sim 8000~{\rm K}$
and 17,000~K, which measures the strength of the Balmer jump resulting
from a competition between bound-free atomic hydrogen and free-free
H$^-$ opacities (see \citealt{weidemann71},
  \citealt{shipman80}). We will attempt later to exploit this
particular sensitivity to $\logg$ (see Section \ref{sect:logg_phot}).

\subsection{Photometric Technique}\label{sect:techphot}

Atmospheric parameters, $\Te$ and $\logg$, of white dwarf stars can be
measured with the photometric technique described in \citet{BRL97},
where photometric measurements are converted into spectral energy
distributions, which are then compared with those predicted from model
atmosphere calculations.  Although usually applied in the context of
cool degenerates when the optical spectra become almost completely
featureless, we will apply this method to all objects in our sample
and will attempt to evaluate its validity over the entire temperature
range covered by the SDSS. In the case of the SDSS $ugriz$ photometry,
the magnitude system is defined in terms of AB$_{95}$ magnitudes, where
we apply a correction to the $u$, $i$, and $z$ bands of $-0.040$,
$+0.015$, and $+0.030$, respectively, to account for the
transformation from the SDSS to the AB$_{95}$ magnitude system, as
explained in \citet{eisenstein06}. We first transform every magnitude
$m_\nu$ into an average flux $f^{m}_{\nu}$ using the equation

\begin{equation}
m_\nu=-2.5\log f^{m}_{\nu} - 48.60, \label{eq:1}
\end{equation}

\noindent where

\begin{equation}
f^{m}_{\nu}=\frac{\int f_{\nu}S_m(\nu)\,d\log\nu}{\int S_m(\nu)\,d\log\nu}, \label{eq:2}
\end{equation}

\noindent and where $f_\nu$ is the monochromatic flux from the star
received at Earth, and $S_m(\nu)$ is the total system response
including the atmospheric transmission corresponding to an air mass of
1.3 and mirror reflectance, as well as the detector quantum efficiency
\citep{HB06}. A set of transmission curves for the SDSS filters
measured by Jim Gunn in 2001 is available on the survey's
website\footnote{http://www.sdss3.org/instruments/camera.php}. A more
recent estimate of these curves has been published by
\citet{Doi2010}, and this will be the set of transmission curves used
here; a comparison of the photometric temperatures obtained with these two sets of
filters is presented in Section \ref{sect:filters}. Since the
observed fluxes correspond to averages over given bandpasses, the
monochromatic fluxes from the model atmospheres need to be converted
into average fluxes as well, $H^{m}_{\nu}$, which can done by substituting $f_{\nu}$
in Equation \ref{eq:2} with the monochromatic Eddington flux
$H_{\nu}$. We can then relate the average observed fluxes
$f^{m}_{\nu}$ and the average model fluxes $H^{m}_{\nu}$ --- which
depend on $\Te$ and $\logg$ --- by the equation

\begin{equation}
f^{m}_{\nu} = 4\pi(R/D)^2H^{m}_{\nu} \label{eq:3}
\end{equation}

\noindent where $R/D$ defines the ratio of the radius of the star to
its distance from Earth. In the above equation, the radius $R$ is
obtained from the $\logg$ value by using evolutionary models
similar to those described in \citet{FBB01} but with C/O cores,
$q({\rm He})\equiv \log M_{\rm He}/M_{\star}=10^{-2}$, and $q({\rm
  H})=10^{-4}$, which are representative of hydrogen-atmosphere white
dwarfs. We then minimize the $\chi^2$ value defined in terms of the
difference between observed and model fluxes over all bandpasses,
properly weighted by the photometric uncertainties. Our minimization
procedure relies on the nonlinear least-squares method of
Levenberg-Marquardt \citep{press86}, which is based on a steepest
decent method. In principle, for stars with known trigonometric
parallax measurements, the distance $D$ in Equation \ref{eq:3} can be
obtained directly, in which case the minimization procedure yields the
effective temperature and the radius of the star. However, for all
white dwarfs in the SDSS, for which parallaxes are not available, we
will simply assume a value of $\logg=8.0$, although we will also
experiment with spectroscopic $\logg$ values (see Section
\ref{sect:log_g}). In these cases, $\Te$ and the solid angle $\pi
(R/D)^2$ are considered free parameters, and the uncertainties of both
parameters are obtained directly from the covariance matrix of the
fit.

To measure the atmospheric parameters from photometry, we rely on two
different sets of model atmospheres and synthetic spectra.  For
effective temperatures below 30,000 K, synthetic spectra are
calculated using the LTE approximation and the ML2/$\alpha$=0.7
version of the mixing-length theory to treat the atmospheric
convection, which becomes important below $\Te=15,000~{\rm K}$, using
the model atmosphere code described at length in \citet{TB09} and
references therein. For $\Te>30,000~{\rm K}$, NLTE effects are taken
into account using TLUSTY \citep{TLUSTY}. Combining these two grids,
we obtain model spectra for effective temperatures ranging from
$\Te=1500~{\rm K}$ to $\Te=120,000~{\rm K}$ and for surface gravities
between $6.0 \leq \logg \leq 9.5$. These models
also include the improved Stark broadening profiles from \citet{TB09}.

\subsection{Selection of the Sample}

Since our goal is to compare the atmospheric parameters obtained from
photometric and spectroscopic data, we want to retain only the best
data available for each set. It was mentioned earlier that the
photometric sample was limited to DA white dwarfs with a
signal-to-noise ratio above 25 in the $g$ band and
  $\sigma_{m_\nu}<0.1$ in all bandpasses (see Section
  \ref{sect:phot_sample}). These criteria alone are not sufficient to
eliminate all bad data from our sample, however, as illustrated in
Figure \ref{fig:fits_photo_exemple} where we show representative
photometric fits as a function of increasing reduced $\chi^2$
  values (i.e., divided by number of degrees of freedom). While the
fits displayed in the left panels are excellent, those in the right
panels are more problematic. Since the photometric technique is based
on a $\chi^2$ minimization approach, we can use these results to
define a critical value $\chi^2_{\rm crit}$ above which the $ugriz$
data are considered unreliable and are excluded from our photometric
sample.

Figure \ref{fig:chisq} shows the reduced $\chi^2$ distribution
obtained from our fits using the photometric technique. This
distribution reveals that for most of the objects, the photometric
energy distributions are well reproduced (low $\chi^2$
values). However, for several stars, the observed magnitudes cannot be
fitted properly by the photometric technique (high $\chi^2$
values). Since we do not want to consider these objects in our
comparative analysis, we define a value for $\chi^2_{\rm crit}$ above
which the photometric fits will not be considered accurate enough for
our purposes. By examining the results of our photometric fits and
corresponding $\chi^2$ values, we have arbitrarily determined that by
setting $\chi^2_{\rm crit}=3$, most bad fits were successfully
excluded from the sample. According to this criterion, all fits
  displayed in the right panels of Figure \ref{fig:fits_photo_exemple}
  would be excluded.

This additional reduced $\chi^2$ criterion ensures that we now
have a clean photometric sample. In some cases, however, this
criterion may lead to the exclusion of good data sets. Indeed, since
we assumed $\logg=8.0$ for all objects in our sample, white dwarfs
with effective temperatures between $\Te \sim 8000~{\rm K}$ and $\sim
17,000~{\rm K}$, for which the $u-g$ color index is sensitive to
surface gravity (see Figure \ref{fig:umg_gmr}), may be excluded from
our sample if their $\logg$ value differs significantly from
$8.0$. For instance, the object 042017.86+052735.8 displayed in Figure
\ref{fig:fits_photo_exemple} has a {\it spectroscopic} $\logg$ value
of $8.362$, and forcing $\logg=8.0$ thus leads to a bad photometric
fit and corresponding large value of $\chi_{\rm red}^2\sim10$. One
solution to this problem would be to use only the $griz$ photometry,
thus avoiding the $u$ bandpass which is $\logg$ sensitive. However, as
discussed in Section \ref{sect:result_photo}, the $u$-band photometry
is important for the analysis of hot white dwarfs ($\Te \gta
20,000~{\rm K}$), and we want to use a consistent approach for all
stars, regardless of their effective temperature.  Another solution
would be to rely on spectroscopic $\logg$ values since all white
dwarfs in our photometric sample have a measured SDSS spectrum. But
since the goal of our study is to compare the atmospheric parameters
measured {\it independently} from the photometric and spectroscopic
techniques, we prefer to assume $\logg=8.0$ throughout. Considering
these facts, we will retain all objects in our analysis, but will
display those with $\chi_{\rm red}^2>\chi^2_{\rm crit}$ with a
different color symbol.

\subsection{Selected Results}\label{sect:result_photo}

\subsubsection{Transmission Curves}\label{sect:filters}

The SDSS filters were designed to collect fluxes in the ultraviolet
($u$), green ($g$), red ($r$), near-infrared ($i$), and far infrared
($z$). For the $g$, $r$, and $i$ filters, the red wavelength cutoff is
achieved by applying an interference coating. These layers dehydrated
when they were placed in the vacuum of the camera, changing the
refractive index of the interference coating. The dehydration caused
the red edges of these three filters to be shifted blueward
\citep{Fan2001}. However, it seems the $u$ filter was the one that has
changed the most with time \citep{Doi2010}. The $u$ filter has a
natural red leak around 7100
\AA \footnote{http://cas.sdss.org/dr7/en/help/docs/algorithm.asp?key=photometry},
which is usually suppressed by the application of an interference
coating. The refractive index of this interference coating also
changed due to dehydration when the filter was placed in the vacuum of
the camera, so the natural red leak is not completely suppressed. We
explore the effects of this red leak on the results of the photometric
temperatures in Section \ref{sect:u_band}.

It was mentioned in Section \ref{sect:techphot} that there are two
sets of transmission curves for the $ugriz$ filters. The first one,
measured by Jim Gunn in 2001, is available on the SDSS website. It is
important to point out that these filter curves do not include the
complete system response from atmosphere to
detector\footnote{http://www.sdss3.org/instruments/camera.php}. A more
recent estimate of these transmission curves, using more data points
and a larger time baseline, was published in \citet{Doi2010}. Both
sets of transmission curves are shown in Figure \ref{fig:DoivsGunn};
note that each curve has been normalized to unity for easier
comparison. Since the curves measured by Gunn in 2001 were used in
several investigations prior to the publication of Doi et al., we
wanted to evaluate the effect of the filter transmission curves on the
photometric results. Figure \ref{fig:comp_DoivsGunn} presents a
comparison of effective temperatures obtained using both sets of
transmission curves. The results clearly show that the photometric
results are not affected by the particular choice of transmission
curves. Since the Doi et al.~curves are more recent and use more data
points, we will use those in the remainder of our analysis.

\subsubsection{The $u$-band}\label{sect:u_band}

As mentioned above, the $u$ filter has a natural red leak around
6000-8000 \AA, which is supposed to be suppressed by an interference
coating \citep{Doi2010}. The dehydration of this coating shifted the
wavelength cutoff blueward, thus the problem is only partially
corrected and a small leak still remains near 7700 \AA. Since we want
to compare atmospheric parameters determined from photometry and
spectroscopy, we have to make sure that the $u$-band red leak does not
affect the results of the photometric technique. To achieve this, we
can compare effective temperatures obtained from $ugriz$ photometry
with those obtained by ignoring the $u$-band photometry. These results
are shown in Figure \ref{fig:comp_ugrizvsgriz}.

Cooler stars have an important flux contribution in the red region of
the electromagnetic spectrum, and the red leak could potentially
affect their photometric temperatures more significantly. But as shown
in Figure \ref{fig:comp_ugrizvsgriz}, the effective temperatures
obtained using $ugriz$ or only $griz$ are very similar for
$\Te<20,000~{\rm K}$. This suggests that the red leak does not affect
the photometric results significantly for cool white dwarfs. At higher
temperatures, however, the scatter in the distribution becomes more
important. Since hot white dwarfs emit very little flux in the red
portion of the electromagnetic spectrum, it is unlikely that the
scatter has anything to do with the $u$ filter red leak. A more likely
explanation is that any photometry in the optical becomes less
sensitive to $\Te$ for hotter stars, as the energy distribution falls
into the Rayleigh-Jeans regime, and one needs to push the photometry
further into the ultraviolet.  Hence the increased scatter observed in
Figure \ref{fig:comp_ugrizvsgriz} above $\Te\sim20,000$~K is certainly
due to larger uncertainties in photometric temperatures based on
$griz$ photometry only, and not to the $u$-band red leak, and these
results stress the importance of including $u$-band photometry to
estimate the effective temperature of hotter white dwarfs.

\subsubsection{Effects of the Surface Gravity}\label{sect:log_g}

Since no parallax measurement is available for any of the SDSS white
dwarfs in our sample, we simply assume a value of $\logg=8.0$ to
estimate the photometric temperatures.  However, since surface gravities for white
dwarfs are found in a wide range of values, $6.5 \lesssim \logg \lesssim
9.5$, our assumption of $\logg=8.0$ could have an effect
on the photometric temperatures, especially in the range
$8000~{\rm K}<\Te<17,000~{\rm K}$ where the Balmer jump is 
sensitive to surface gravity (see Figure \ref{fig:umg_gmr}). As all
white dwarfs in our photometric sample have been spectroscopically
identified in the SDSS, they also have a measured spectrum. Therefore,
we can apply the spectroscopic technique (see Section
\ref{sect:tech_spect}) to measure their surface gravities (properly
corrected for 3D hydrodynamical effects --- see Section
\ref{sect:3D}), and then use these spectroscopic $\logg$ 
values when fitting the photometry. We can then compare the corresponding
photometric temperatures with those obtained under the assumption of
$\logg=8.0$. The results of this experiment are displayed in Figure
\ref{fig:comp_gspectro8}.

 Our results indicate that the assumption of $\logg=8.0$ does not
  affect the photometric temperatures significantly, even in the
  region where the $u-g$ color index is particularly sensitive to
  surface gravity ($8000~{\rm K}<\Te<17,000~{\rm K}$). This might be a
  suprising result at first given the strong $\logg$ dependence
  illustrated in Figure \ref{fig:umg_gmr} in this temperature
  range. And indeed, the photometric temperature of a $\Te=14,700$~K
  object at $\logg=8.0$ drops to 14,000~K if we assume a value of
  $\logg=8.5$ instead, i.e.~a 700~K temperature difference. However, a
  closer inspection of the results shown in Figure
  \ref{fig:comp_gspectro8} between 9000~K and 17,000~K reveals that
  most temperature differences lie well within 500~K, and that only 5
  objects exceed this value, mainly because the surface gravity
  distribution is so strongly peaked around $\logg=8.0$
  ($\sigma_{\logg}\sim0.2$ in this temperature range; see Figure
  \ref{fig:corr_spectro_SDSS} below). Actually, one of these 5 objects
  is 042017.86+052735.8, already discussed above (see Figure
  \ref{fig:fits_photo_exemple}), with a spectroscopic value of
  $\logg=8.362$ significantly above average.  This indicates that it
  would be more accurate to rely on spectroscopic $\logg$ values, but
again, since our goal is to compare the temperatures derived
independently from the photometric and spectroscopic techniques, we
will assume $\logg=8.0$ for the remainder of this analysis.

\subsection{Photometric Calibration}\label{sect:phot_calib}

Our next step is to ensure that the $ugriz$ photometry is properly
calibrated. It is well known that the $ugriz$ photometric system is
not entirely consistent with the AB$_{95}$ system, and that small
zeropoint offsets exist. Theses offsets in the $u$, $i$, and $z$ bands
can be compensated by applying the appropriate corrections from
\citet{eisenstein06}, as discussed in Section \ref{sect:techphot}. In
this section, we attempt to validate the overall calibration of the
$ugriz$ photometric system.  One way to achieve this goal was
presented by \citet{HB06} where the standard star Vega and four
fundamental HST white dwarfs were used to test the calibration of the
$UBVRI$, Str\"omgren, 2MASS, and $ugriz$ photometric systems. By
combining the measured spectrum and the proper set of transmission
curves, they obtained computed magnitudes for each system, which were
then compared to the observed photometry. The comparison was also
extended to a set of 107 DA stars with $ugriz$ photometry and
spectroscopic values of $\Te$ and $\logg$ available (see their Table
14). A disadvantage of this approach is that it relies on
spectroscopic data as well as on the spectroscopic technique, while we
would prefer a method that is completely independent from
spectroscopy. We thus propose below a different method.

The photometric technique discussed above relies on a $\chi^2$
minimization procedure to find the model energy distribution that best
reproduces the observed photometry. We can therefore compare the {\it
  differences} between the observed $ugriz$ data and the theoretical
energy distribution predicted by the photometric technique. To avoid
any bias, we do not restrain the sample to objects for which $\chi_{\rm red}^2 <
\chi^2_{\rm crit}$. We do, however, consider only objects with $\Te
\leq 20,000~{\rm K}$ since the photometric technique becomes less
sensitive to temperature above this range; we also apply the SDSS to
AB$_{95}$ zeropoint corrections from \citet{eisenstein06}.  The
results for our photometric sample are displayed in Figure
\ref{fig:histo}, where we show histogram distributions between
observed (obs) and theoretical (th) magnitudes for each individual
bandpass of the $ugriz$ system.  Our results show that all
histograms appear symmetrical and well centered on $m_{\nu,{\rm
    obs}}-m_{\nu,{\rm th}}=0.0$, which indicate that the $ugriz$
photometric system is properly calibrated, at least in a relative
sense. Indeed, if the $g$ photometry, say, was not well calibrated, it
would be systematically lower (or larger) than the predicted
photometry, and the corresponding histogram would not be
centered. Since the photometric technique fits the complete energy
distribution, and not each photometric point individually, the $u$ and
$r$ distributions would be shifted as well, but in the opposite
direction. This is nicely illustrated in Figure
\ref{fig:histo_corrections}, where the SDSS to AB$_{95}$ corrections
have not been applied. Except for the $i$-band, all histograms are not
centered, a result that demonstrates the necessity to apply these
zeropoint corrections.

Despite these reassuring results, there could still be a calibration
issue that depends on the observed magnitude. For example, if
$m_{\nu,{\rm obs}}>m_{\nu,{\rm th}}$ for bright objects and
$m_{\nu,{\rm obs}}<m_{\nu,{\rm th}}$ for faint objects, the histograms
displayed above would still be centered on average. To investigate
this possibility, we illustrate in Figure \ref{fig:dispersion} the
same magnitude differences, $m_{\nu,{\rm obs}}-m_{\nu,{\rm th}}$, but
this time as a function of the observed magnitude. As can be seen, all
distributions are centered regardless of the magnitude. Moreover,
except for the $z$-band, the scatter in the distributions remains
somewhat constant with the observed magnitude. For the $z$-band, the
dispersion becomes more important with larger $z_{\rm obs}$. This can
also be observed in the histograms from Figure \ref{fig:histo} where
$\sigma_z=0.041$, while $\sigma_{m_\nu} \sim 0.02-0.03$ for the other
bands. The dispersion affects mostly objects that are very faint in
the mid-infrared region ($z>18.0$). Since the SDSS camera images 1.5
deg$^2$ at once, the exposure time is the same for every object and
this might not be sufficient enough to ensure a good signal-to-noise
ratio in the $z$-band. This is also reflected in the photometric
uncertainties, where we obtain an average of
$\langle\sigma_z\rangle=0.031$ for the overall sample, compared to
$\sim$0.02 for the other bands.

In the same context, we would like to point out that the mean
uncertainties for each photometric band --- $\langle\sigma_u\rangle=0.021$,
$\langle\sigma_g\rangle=0.019$, $\langle\sigma_r\rangle = 0.016$,
$\langle\sigma_i\rangle = 0.018$, and $\langle\sigma_z\rangle = 0.031$ ---
are somewhat smaller than the standard deviations of the
distributions observed in Figure \ref{fig:histo} --- $\sigma_u=0.032$,
$\sigma_g=0.027$, $\sigma_r=0.018$, $\sigma_i=0.022$, $\sigma_z=0.041$,
which suggests that the quoted photometric uncertainties of the $ugriz$
data might be slightly underestimated.

\subsection{Interstellar Reddening}\label{sect:reddening}

Due to the nature of the Sloan Digital Sky Survey, white dwarf stars
in our photometric sample are particularly faint ($16<g< 20$) --- see
Figure \ref{fig:dispersion}. Consequently, some white dwarfs in this
sample may be quite distant, and their magnitudes are thus likely to
be affected by interstellar reddening. We can estimate the distance to
each star in our sample by using the photometric technique, which
yields the value of the solid angle $\pi (R/D)^2$, and thus the
distance $D$ for a stellar radius $R$ corresponding to our assumed
value of $\logg=8$ at the photometric temperature (derived from our
evolutionary models). Since this assumption on $\logg$ directly
affects our distances, we can improve upon these estimates by relying
instead on the spectroscopic $\logg$ values described in the next
section. These {\it photometric distances} for our SDSS sample are
shown in Figure \ref{fig:distances_sdss}. As this figure shows, many
stars in this sample are indeed found at large distances ($>100$ pc),
implying that their magnitudes are most likely affected by
interstellar extinction. We thus modified our photometric technique to
deredden the observed magnitudes. Our procedure is identical to that
described in \citet{Tremblay2011}, based on the parameterization of
\citet{harris06} for the amount of reddening as a function of
distance.  This procedure works in an iterative fashion, using the
distance of the star found from the previous iteration. Interstellar
absorption is assumed to be negligible for stars with distances less
than 100 pc, and maximum for stars with distances $|z| > 250$ pc from
the Galactic plane. The absorption is assumed to vary linearly along
the line of sight between these two regimes.

Figure \ref{fig:red_vs_dered} compares the effective temperatures
obtained from undereddened and dereddened $ugriz$ magnitudes. As
expected, temperatures obtained from dereddened magnitudes are
systematically higher than those measured with undereddened
photometry. The effect of interstellar reddening is
particularly important for hotter stars ($\Te>12,000$~K). Since most
stars in the SDSS are faint, the hotter white dwarfs, which are
intrinsically more luminous, are likely to be more distant than their
cooler siblings, and thus more affected by reddening. We also note in
Figure \ref{fig:red_vs_dered} an increase in dispersion at higher
temperatures, which probably reflects a spread in distances, and thus
in the corresponding amount of reddening.

As Figure \ref{fig:red_vs_dered} clearly illustrates, the photometric
temperatures are particularly sensitive to interstellar extinction,
especially for hot white dwarfs, and all $ugriz$ magnitudes will be
dereddened in the remainder of our analysis, unless otherwise specified. We
also keep in mind, however, that our procedure for taking into account
the effects of interstellar reddening remains approximate.

\section{Spectroscopic Analysis}

\subsection{Spectroscopic Samples}

Since the DR7 White Dwarf Catalog \citep{DR7} contains only
spectroscopically identified white dwarfs, all stars in our
photometric sample also have a measured spectrum in the
SDSS database. Therefore, our spectroscopic sample is composed of the same
stars as the photometric sample. All spectra for this sample have been
acquired from the SDSS Data Archive Server\footnote{http
  ://das.sdss.org/spectro/} (DAS), which contains the data up to DR7,
inclusively. The spectra were reduced with the DR7 data reduction
algorithm\footnote{http://www.sdss.org/dr7/algorithms/index.html};
each spectrum has a spectral coverage from 3800 \AA~to 9200 \AA~with a
resolution of 3 \AA~(FWHM).

We also use for comparison the spectroscopic sample from
\citet{Gianninas2011}. This sample contains 1150 bright ($V \leq
17.5$) DA white dwarfs drawn from the online version of the Villanova
White Dwarf Catalog of
\citet{mccook99}. These spectra have a high signal-to-noise ratio
($S/N \sim 70$) and were acquired with different instruments over a
time period of about 20 years, so the spectral coverage and 
spectral resolution differ from one spectrum to another (from
3~\AA\ to 9 \AA\ FWHM); for more information on data acquisition, see
Section 2 of \citet{Gianninas2011}.

\subsection{Spectroscopic Technique}\label{sect:tech_spect}

The best method for measuring the atmospheric parameters of DA stars
using spectroscopic data was first discussed in detail by
\citet{BSL92}. This so-called spectroscopic technique was then
improved by \citet{BSW95}, and more recently by \citet{LBH05}. This
technique allows us to determine the effective temperature and the
surface gravity of a white dwarf by comparing its observed spectrum to
a grid of model spectra. The first step is to normalize the flux from
each individual line to a continuum set to unity at a fixed distance
from the line center, for both observed and model spectra. Observed
and synthetic spectra are then compared in terms of line shapes
only. There are two approaches to define the spectrum continuum. The
first one is used when the star's effective temperature is in the
interval $9000~{\rm K}<\Te<16,000~{\rm K}$ where the Balmer lines are
strong. The normalization is then performed using a sum of
pseudo-gaussian profiles, which prove to be a good approximation for
the observed Balmer lines. If the star lies outside of this
temperature range, Balmer lines become weak and the continuum between
those lines is essentially linear, therefore pseudo-gaussian profiles
cannot be used as easily. We rely instead on model spectra to
reproduce the overall spectrum. In this case, we include a wavelength
shift and several order terms in $\lambda$, up to $\lambda^6$, using
the nonlinear least-square method of Levenberg-Marquardt
\citep{press86}. At this point, we have a smooth model fit, but the
values of $\Te$ and $\logg$ obtained in this manner are meaningless
since too many fitting parameters are used. Now that every line is
normalized, we can proceed to determine the values of $\Te$ and
$\logg$ using our grid of model spectra, convolved with the
appropriate instrumental gaussian profile (3, 6, or 9 \AA, depending
on the resolution of the observed data), and the same fitting method.
When the effective temperature of the white dwarf is close to the
region where the equivalent widths of the Balmer lines reach their
maximum ($\Te\sim 13,500$~K) two solutions are possible, one on each
side of the maximum. Here we take advantage of the available $ugriz$
photometry and adopt the photometric temperature obtained previously
as the starting point of this iterative process.

Sample fits using the spectroscopic technique are displayed in Figure
\ref{fig:ex_spectro}. As illustrated here, some of the SDSS spectra in
our sample were found to be problematic (see, e.g., bottom fits in
Figure \ref{fig:ex_spectro}) and these have been removed from our
sample.

\subsection{3D Hydrodynamical Corrections}\label{sect:3D}

Even though the spectroscopic technique is arguably the most {\it
  precise} technique for measuring the atmospheric parameters of DA
stars, $\Te$ and $\logg$, we still need to determine whether the resulting
parameters are also {\it accurate}.  One way to accomplish this
is to compare the atmospheric parameter distributions with the predictions of
evolutionary tracks at constant masses. Such a comparison for the SDSS and
the Gianninas spectroscopic samples is shown in the top panels of
Figures \ref{fig:corr_spectro_SDSS} and
\ref{fig:corr_spectro_Gianninas}, respectively, together with a single mass
evolutionary track corresponding to the median mass of each sample (see below).

In both distributions, it is obvious that surface gravities are
overestimated at low effective temperatures ($\Te<13,000~{\rm
  K}$). This corresponds to the well documented high-$\logg$ problem
(see, e.g., \citealt{Tremblay2010} and references therein). Many
scenarios have been proposed in the past to account for this problem,
the most popular of which involves convective mixing.  Since a
hydrogen-atmosphere white dwarf becomes convective below $\Te \sim
15,000~{\rm K}$, a significant amount of helium can
be convectively mixed with the outer hydrogen atmosphere if the
hydrogen layer is thin enough
\citep{Koester1976,Vauclair1977,Dantona1979}. Helium would remain
practically invisible at these temperatures, but the spectroscopic
$\logg$ values would appear higher than average when measured with
pure hydrogen models \citep{BWF91}. However, no trace of atmospheric
helium has been reported in high dispersion, high S/N spectra
\citep{Tremblay2010}, ruling out this scenario as a possible
explanation for the high-$\logg$ problem. It is now commonly accepted
that this overestimation is caused instead by the use of the
mixing-length theory (MLT) to treat convective energy transport in 1D
model atmospheres, and that the problem can be solved with the use of
3D hydrodynamical model atmospheres
\citep{TremblayI}. \citet{TremblayIV} recently calculated a new grid
of 3D model atmospheres and published correction functions (in both
$\Te$ and $\logg$) that can be applied to atmospheric parameters
obtained from 1D/MLT models. These correction functions, reproduced
here in Figure \ref{fig:functions_Tremblay}, indicate that below
$\Te\sim15,000$~K, i.e.~when atmospheric convection becomes important,
surface gravities obtained from 1D models are overestimated with
respect to 3D models, and that the maximum corrections occur near
$\sim$10,000~K regardless of the surface gravity of the star.

The atmospheric parameters for the SDSS and the Gianninas
spectroscopic samples, corrected for 3D hydrodynamical effects, are
shown in the bottom panels of Figures \ref{fig:corr_spectro_SDSS} and
\ref{fig:corr_spectro_Gianninas}, respectively.  The high $\logg$
problem has now vanished, and both the SDSS and the Gianninas $\logg$
distributions follow the evolutionary track well below
$\Te\sim13,000$~K. However, a closer inspection reveals that the
atmospheric parameters for both samples are not fully corrected, and
that surface gravities are still overestimated near
$\Te\sim12,000$~K. The cause of this slight overestimation is not fully
understood, but \citet{TremblayIV} suggest that this might be related
to the opacity sources or the equation-of-state in the model
atmospheres.

Another way to investigate the high-$\logg$ problem is to compare the
mass distributions of hot (radiative atmosphere) and cool (convective
atmosphere) DA stars (unlike the surface gravity, the mass of a white
dwarf remains constant with time). The mass distributions for white
dwarfs in the SDSS and the Gianninas samples are displayed in Figures
\ref{fig:histo_mass_SDSS} and \ref{fig:histo_mass_Gianninas},
respectively, for both the 1D/MLT models (left panels) and with the 3D
corrections applied (right panels); the individual contributions for
cool ($\Te<13,000$~K) and hot ($\Te>13,000~{\rm K}$) white dwarfs are
also shown. As observed in the left panels of these figures, the
uncorrected mass distributions for hotter objects peak around $M \sim
0.6$ \msun\ for both samples, while they peak around $M \sim 0.7$
\msun\ for cooler objects. When the 3D corrections are applied,
however, both distributions for hot and cool white dwarfs peak at $M
\sim 0.6$ \msun. The mean mass of cool white dwarfs is still
$\sim$$0.02-0.03$ \msun\ larger than that of hotter objects, but this can
probably be explained by the fact that the fraction of massive white
dwarfs appears higher at low temperatures. We also notice that the
mean mass of the Gianninas sample is about $\sim$0.03 \msun\ larger
than the SDSS sample, most likely due to a residual problem with the
flux calibration of the SDSS spectroscopic data (see also \citealt{Tremblay2011},
\citealt{Gianninas2011}, and next subsection).

\subsection{Spectroscopic Calibration}\label{sect:spec_calib}

As for the photometric sample, it is important to make sure that the
spectroscopic data are properly calibrated. One way to accomplish this
is to compare the behavior of the atmospheric parameter distributions
(corrected for 3D hydrodynamical effects) for the SDSS and Gianninas
samples, displayed in the bottom panels of Figures
\ref{fig:corr_spectro_SDSS} and \ref{fig:corr_spectro_Gianninas},
respectively. A first discrepancy between these two distributions can
be observed around $\Te\sim14,000~{\rm K}$ where there is a small but
significant accumulation of objects for the Gianninas sample, while
there appears to be a small deficit for the SDSS sample in the same
temperature range. This corresponds precisely to the temperature 
where the strength of the hydrogen Balmer lines reach their maximum.
Since the spectroscopic technique relies on the strength of the
hydrogen lines, such accumulations or gaps may appear in the
temperature distributions if the lines in the model spectra are
predicted too weak or too strong in this temperature range (see also
Figure 3 of \citealt{bergeron95} for a similar result).  The fact that
there is an accumulation of objects in that region for the Gianninas
sample indicates that the models predict weaker lines than what is
observed, while the opposite occurs for the SDSS sample.  This
difference in behavior suggests that at least one of our spectroscopic
samples has a calibration issue.

Another discrepancy is visible when we compare both distributions to
the evolutionary tracks at a constant mass, as depicted in Figures
\ref{fig:corr_spectro_SDSS} and \ref{fig:corr_spectro_Gianninas}.
While the Gianninas distribution follows the constant (median) mass
evolutionary track through the entire temperature range displayed
here, the surface gravities for the SDSS objects fall below the track
above $\sim$16,000~K, and slightly above the track at lower
temperatures, indicating that the SDSS spectroscopic sample may have a
small calibration issue. This problem is well known and the SDSS data
reduction algorithm has been improved several times to correct for
this problem. A description of these improvements is available on the
SDSS
website\footnote{http://www.sdss.org/dr7/algorithms/dataProcessing.html}. The
spectroscopic data used here have been reduced with the DR7 version of
the algorithm described in \citet{Algo}. It seems that, despite these
improvements, the SDSS data reduction algorithm is still not perfect.

Finally, since the Gianninas spectroscopic sample covers more or less
the entire sky, some of the white dwarfs lie in the SDSS field, and
thus also have a measured SDSS spectrum. The method used to recover
the spectra in common between both samples is described in the next
section.  Using this method, we identified 200 of the Gianninas white
dwarfs that also have a measured spectrum in the SDSS. Figure
\ref{fig:comp_spec} presents the comparison of effective temperatures
obtained from the SDSS and the Gianninas spectroscopic data. For
$\Te<14,000~{\rm K}$, SDSS spectra yield effective temperatures
slightly lower than those obtained with the Gianninas spectra. For
$\Te>14,000~{\rm K}$, however, the trend is more significant and in
the opposite direction, with the SDSS temperatures being $\sim$1100~K
higher, on average, than the Gianninas temperatures.  Again, we
believe that the SDSS spectroscopic data suffer from a small
calibration problem, and we will keep this in mind when comparing
spectroscopic and photometric temperatures.

\section{Comparison of Atmospheric Parameters}\label{sect:comp_param}

The goal of our study is to compare the atmospheric parameters (in
particular the effective temperatures) obtained from photometric and
spectroscopic observations. We will first consider the SDSS sample, for which
$ugriz$ photometry and optical spectroscopy are available for the same
objects. We showed in Section \ref{sect:reddening} that white dwarfs
in the SDSS are generally found at large distances, and thus that the
observed magnitudes may be significantly affected by interstellar
reddening. We also determined in Section \ref{sect:3D} that 3D
hydrodynamical corrections needed to be applied to the atmospheric
parameters determined from the spectroscopic technique. In the
remainder of our analysis, we thus systematically deredden the $ugriz$
photometry following the procedure outlined in Section
\ref{sect:reddening}, and apply the 3D correction functions, unless
otherwise specified.

The comparison of effective temperatures obtained from photometry and
spectroscopy for the SDSS sample is presented in Figure
\ref{fig:comp_phot_SDSS_rougissement_g2}. Overall, the photometric and
spectroscopic temperatures agree surprisingly well, even at high
temperatures where the energy distribution becomes insensitive to
variations in $\Te$ in the optical (see also Figure
\ref{fig:umg_gmr}). This lack of sensitivity of the $ugriz$ photometry
to effective temperature is most likely responsible for the increased
dispersion observed above $\Te\sim30,000~{\rm K}$.  The dispersion
also becomes important near $\Te\sim14,000~{\rm K}$ in the region
where the strength of the hydrogen lines reach their maximum. As
discussed in Section \ref{sect:spec_calib}, our models predict
stronger lines than those observed in the SDSS spectra, resulting in
the large dispersion observed in Figure
\ref{fig:comp_phot_SDSS_rougissement_g2} in this particular
temperature range.

Despite the overall agreement between spectroscopic and photometric
temperatures, we can observe a small but significant temperature offset above
$\Te\sim14,000~{\rm K}$, where spectroscopic temperatures appear
systematically higher than those determined from photometry (about $\sim$630~K
on average between 15,000~K and 40,000~K). Since
the magnitudes have been corrected for interstellar reddening, and
that this effect is particularly important at higher
temperatures (see Figure \ref{fig:red_vs_dered}), we need to test whether
our correction procedure is responsible for the observed temperature
offset. The importance of interstellar reddening is illustrated in
Figure \ref{fig:comp_spectro_photo_log} where we show the same
comparison between photometric and spectroscopic temperatures, but by
using undereddened magnitudes. The temperature discrepancy becomes
even more significant, especially at high effective temperatures, as
expected. If anything, we can conclude that our procedure for taking
into account the presence of interstellar reddening works fairly well.

We already discussed that the SDSS spectroscopic data suffer from a
small calibration problem, and it is thus conceivable that this
problem might be the origin of the temperature offset observed in
Figure \ref{fig:comp_phot_SDSS_rougissement_g2}. Indeed, we already
showed in Figure \ref{fig:comp_spec} that the effective temperatures
above 14,000~K based on SDSS spectra appear {\it overestimated} with
respect to those obtained from the Gianninas spectra. We can test this
hypothesis by performing a similar experiment using the Gianninas
spectroscopic sample. The first step is to search the SDSS database
for white dwarfs in the Gianninas sample with measured $ugriz$
photometry. Here we restrain our search to objects cooler than
$\Te\sim 40,000~{\rm K}$ where the photometric technique is most
sensitive. Since the SDSS is not an all-sky survey, not all stars in
the Gianninas sample have measured $ugriz$ photometry. To determine if
an object is in the SDSS field, we first use the Simbad
database\footnote{http://simbad.u-strasbg.fr/simbad/} to obtain the
positions of the stars in our sample, which are then fed into the
SDSS Coverage Check
tool\footnote{http://dr10.mirror.sdss3.org/coverageCheck/search}.  We
then compare the finding charts from the Villanova White Dwarf
Catalog\footnote{http://www.astronomy.villanova.edu/WDcatalog/} with
those from the SDSS
website\footnote{http://skyserver.sdss3.org/dr10/en/tools/chart/navi.aspx}
to obtain the position of the stars in the SDSS database, and enter
these positions in the Imaging
Query\footnote{http://skyserver.sdss3.org/dr10/en/tools/search/IQS.aspx}
to retrieve the required $ugriz$ photometric data.  The use of finding
charts was preferred to simply entering the Simbad positions directly
in the Imaging Query, a procedure that resulted in a lot of
mismatches.  Following these steps, we identified 561 white dwarfs in
the Gianninas sample with measured $ugriz$ photometry; we will refer
to this particular sample as the {\it Gianninas subset}. By the same
token, we obtain the corresponding SDSS spectrum, if available, from
the same tool. As previously mentioned in Section
\ref{sect:spec_calib}, 200 of the 561 white dwarfs in the Gianninas
subset also had a measured SDSS spectrum.

The comparison of effective temperatures obtained from photometry and
spectroscopy for the Gianninas subset is presented in the left panel
of Figure \ref{fig:comp_G_SDSS}. For an easier comparison, we
reproduce in the right panel the results obtained from the SDSS sample
on the same scale. Unfortunately, the temperature offset is still
present using the Gianninas subset, and extends to even lower
effective temperatures ($\Te\sim10,000$~K) than with the SDSS
sample. The average difference between spectroscopic and photometric
temperatures for $15,000~{\rm K}<\Te<40,000~{\rm K}$ is about 580~K
for the Gianninas subset, only $\sim$50~K smaller than the difference
observed with the SDSS sample. Note that the effect of interstellar
reddening is almost completely negligible for the Gianninas subset
(not shown here), with the exception of the hottest stars, since most
white dwarfs in this sample are relatively bright and nearby.  Given
that we demonstrated that the $ugriz$ photometry has been properly
calibrated and dereddened, and given that we took into account the
appropriate 3D hydrodynamical corrections in our spectroscopic
analysis, we are left with little explanation to account for the
observed temperature offset. Note that a similar offset was also
reported in the comparison performed by \citet[][see their Figure
  16]{TremblayIV}.

Since the spectroscopic technique is more sensitive than the
photometric technique to the details of model atmosphere
calculations, in particular the line broadening theory, the
occupation probability formalism, etc., we perform a final test using
a different set of model spectra. Here we rely on the model
atmospheres and synthetic spectra described at length in \citet{LBH05}
and references therein. These are based on the Stark profiles from
\citet{Lemke1997}, while our current model spectra rely on the
improved Stark profile calculations from \citet{TB09}, which take into
account nonideal perturbations from protons and electrons ---
described within the occupation probability formalism of \citet{HM88}
--- directly inside the line profile calculations. To compensate for
the neglect of these nonideal effects in the previous Stark broadening
tables, \citet{bergeron93} suggested to include an ad hoc parameter
to {\it mimic} the nonideal effects
in the line profiles, more specifically, by taking twice the value of
the critical electric microfield ($\beta_{\rm crit}$) in the
Hummer-Mihalas theory (see \citealt{TB09} for a full discussion of
this approach).  By doing so, it was found that the internal
consistency between the spectroscopic solutions obtained when
increasing the number of Balmer lines in the fitting procedure was
improved substantially. As also mentioned by Tremblay \& Bergeron,
however, we must emphasize that this is just a quick and dirty way to
simulate the nonideal effects by reducing the line wing opacity, and
this does not imply that Hummer \& Mihalas underestimated the value of the critical
field in any way. The approach described in \citet{TB09} where the
Hummer-Mihalas formalism is included directly into the line profile
calculation, without any modification to $\beta_{\rm crit}$, is more
physically sound by any standard.  

This being said, we redetermined self-consistently the photometric and
spectroscopic effective temperatures using this older generation of
model spectra. The results for the Gianninas subset, which appears to
be better calibrated than the SDSS sample, are shown in Figure
\ref{fig:spec_photo_old_rouge_g}.  First of all, the temperature
offset apparent in Figure \ref{fig:comp_G_SDSS} has been significantly
reduced above $\Te\sim20,000$~K, if not eliminated, but another offset
has now developed between $\Te=13,000~{\rm K}$ and $19,000~{\rm K}$,
where the photometric temperatures now exceed the spectroscopic
values. The average difference between spectroscopic and photometric
temperatures for $15,000~{\rm K}<\Te<40,000~{\rm K}$ is now $-190$~K.
Even though we do not claim that the model spectra described in \citet{LBH05}
are more appropriate than those used here, the results of our
experiment nevertheless suggest that the physics of line
broadening may still require some improvement.

\section{Surface Gravity Determinations using Photometric Data}\label{sect:logg_phot}

As shown in Figure \ref{fig:umg_gmr}, the strength of the Balmer jump,
as measured by the $u-g$ color index, is very sensitive to surface
gravity in a certain range of effective temperature ($8000~{\rm
  K}\lesssim\Te\lesssim 17,000~{\rm K}$). \citet[][see also
    \citealt{shipman80}]{weidemann71} successfully explained this
behavior in terms of a competition between bound-free atomic hydrogen
and free-free H$^-$ opacities. We attempt in this section to exploit
this particular sensitivity to $\logg$ to measure surface gravities
using the photometric technique, an approach similar to that described
by \citet{Koester2009}.

In order to determine the surface gravities for the white dwarfs in
the SDSS sample using photometry, we simply modify the photometric
technique described in Section \ref{sect:techphot} by adding a third
fitting parameter, $\logg$, to the two existing ones --- the effective
temperature $\Te$ and the solid angle $\pi (R/D)^2$. Note that unless
the distance $D$ is known from a trigonometric parallax measurement,
$\logg$ and $\pi (R/D)^2$ are independent parameters.  To achieve a
better convergence of the $\chi^2$ minimization procedure, we obtain a
first estimate of the effective temperature by assuming $\logg=8.0$,
and then we let all three parameters vary. Interstellar reddening is also
taken into account following the same procedure as before.

The atmospheric parameters obtained from this modified photometric
technique are displayed in Figure \ref{fig:logg_phot_dered}; note that
we do not show here objects for which $\chi_{\rm red}^2>\chi^2_{\rm crit}$.
Also, we restrict our analysis to white dwarfs with $8000~{\rm
  K}<\Te<17,000~{\rm K}$ where the sensitivity of the Balmer jump to
surface gravity is most important. Our results reveal that the photometric $\logg$
distribution follows the 0.598 \msun\ evolutionary track quite well,
but the scatter is admittedly significant. The dispersion in $\logg$
has a distinctive butterfly shape, which reaches a minimum around
$\Te=12,000~{\rm K}$ where the sensitivity of the $u-g$ color index to $\logg$
becomes maximal. More interestingly, the high-$\logg$ problem observed
in the top panel of Figures \ref{fig:corr_spectro_SDSS} and
\ref{fig:corr_spectro_Gianninas} is not present in the photometric
$\logg$ distribution, which indicates that the 3D hydrodynamical
effects do not affect the region where the continuum is formed, but
have an effect mostly on the shape of the absorption lines, as
discussed by \citet{TremblayIV}. Consequently,
photometric results remain unaffected by the particular treatment of
convective energy transport at low effective temperatures.

For comparison, we also reproduce in Figure \ref{fig:logg_phot_dered}
the spectroscopic temperatures and surface gravities (corrected for 3D
effects) for the same objects.  It is not surprising to see here that
the dispersion of the spectroscopic distribution is much smaller than
that obtained from photometry, since the spectroscopic technique is
arguably more precise (but not necessarily more
accurate) than any other method for measuring atmospheric parameters
(see, e.g., \citealt{BSL92}).  We explore this more quantitatively in
Figure \ref{fig:histo_logg_dered} where the photometric and
spectroscopic $\logg$ distributions are plotted as histograms.  As
expected from the results shown in Figure \ref{fig:logg_phot_dered},
both distributions peak around $\logg=8.0$, with $\langle\logg_{\rm
  phot}\rangle=7.994$ and $\langle\logg_{\rm spec}\rangle=8.037$, but
the dispersion using the photometric technique ($\sigma_{\rm
  phot}=0.404$) is much more important than that obtained from
spectroscopy ($\sigma_{\rm spec}=0.206$). Nevertheless, our
results indicate that it is still possible to get a reasonable
estimate of the surface gravity using only photometry.

As mentioned above, \citet{Koester2009} used a similar approach to
measure the atmospheric parameters of bright ($g<19$) DA white dwarfs
from the fourth data release (DR4) of the SDSS (see their Figure
2). Unlike our photometric $\logg$ distribution shown in Figure
\ref{fig:logg_phot_dered}, which follows the evolutionary track
extremely well, their photometric $\logg$ distribution still shows a small
high-$\logg$ problem between $\Te\sim12,000$~K and 9000 K, although
less important than that observed in the corresponding spectroscopic $\logg$
distribution. Moreover, their surface gravities are systematically
lower that the mean value of $\logg=8.0$, regardless of effective temperature,
unlike our distribution shown in Figure \ref{fig:logg_phot_dered}.

\section{Summary and Conclusions}

We tested the calibration of the $ugriz$ photometric system by using
hydrogen-line DA white dwarfs from the Data Release 7 of the Sloan
Digital Sky Survey. We first had to ensure that our sample was clean
and contained only objects with good measured $ugriz$ photometry. To
achieve this, we first limited the DA sample to objects with a
signal-to-noise ratio above 25 in the $g$-band and
  $\sigma_{m_\nu}<0.1$ in all bandpasses. Since our photometric
technique is based on a $\chi^2$ minimization method, we then defined
a value $\chi^2_{\rm crit}$ above which the atmospheric parameters
were considered inaccurate. By examining the photometric fits and
their corresponding reduced $\chi^2$ values, we arbitrarily determined
that a critical value of $\chi^2_{\rm crit}=3$ allowed us to exclude
most of the bad data from our photometric sample. In order to test the
photometric calibration, we first verified that the photometric
temperatures were not affected by the red leak in the $u$ filter, or
by the assumption of $\logg=8.0$ in our photometric fits.  We then
compared the observed and predicted magnitudes in each of the five
bandpasses, and we concluded that, when the proper SDSS to AB$_{95}$
corrections are applied, the $ugriz$ photometric system appears well
calibrated.

We then defined two spectroscopic samples: the SDSS sample, which
contains the same white dwarfs as the photometric sample, and the
Gianninas sample, based on the bright ($V \leq 17.5$) DA
white dwarfs from the Villanova White Dwarf Catalog. Using these two
samples, we first explored the well known high-$\log g$ problem at low
effective temperatures, which can be effectively corrected by applying
the 3D correction functions recently published by
\citet{TremblayIV}. We then compared the atmospheric parameters
obtained for the objects in common between the SDSS and Gianninas data
sets, and concluded that the SDSS spectra may still have a small
calibration problem, despite the improvements in their data reduction
algorithm.

The next step in our analysis was to compare the atmospheric
parameters obtained from photometry and spectroscopy. Using the SDSS
photometric and spectroscopic samples, we observed that a systematic
offset appeared above $\Te \sim 14,000~{\rm K}$, where photometric
temperatures are systematically lower than the spectroscopic
values. This temperature offset cannot be fully explained by the
calibration problem of the SDSS spectra since it is also present when
using the Gianninas spectra.  This offset does not appear to be
related to our procedure for taking into account the presence of
interstellar reddening either, since the Gianninas sample, where the offset is
also observed, contains bright, and mostly nearby white dwarfs, which
are not affected by reddening. Hence, the origin of this small discrepancy
still eludes us, but could be related to the physics of the model
spectra used in the spectroscopic method.

Finally, we exploited the sensitivity of the Balmer jump to surface
gravity between $\Te\sim8000$~K and 17,000 K to measure $\logg$ values
using the photometric method. Our results showed that, even if
photometric $\logg$ values are less precise than spectroscopic values,
it is still possible to obtain a good estimate of the $\logg$
distribution using the photometric technique. In particular, we showed
that the high-$\logg$ problem is not observed in the $\logg$ distribution
derived from photometry, confirming that this problem is related to the 
hydrogen line profiles predicted from 1D/MLT model atmospheres.

\acknowledgements We would like to thank A.~Darveau-Bernier for his
earlier contribution to this project. This work was supported in part
by the NSERC Canada and by the Fund FRQ-NT (Qu\'ebec). This research
has made use of the SIMBAD database, operated at CDS, Strasbourg,
France.

\bibliography{biblio}{}
\bibliographystyle{apj}

\clearpage 

\figcaption[f1] {$(u-g,\ g-r)$ two-color diagram for
  hydrogen-atmosphere white dwarfs with $\Te$ indicated in units of
  $10^3$~K, and log $g$ values of 7.0 (0.5) 9.5 (from bottom to
  top). The observed $ugriz$ photometry is taken from
  \citet{DR7}.\label{fig:umg_gmr}}

\figcaption[f2] {Examples of fits using the photometric
  technique. Error bars show the observed $ugriz$ photometry while the
  dots represent the best model fit. The effective temperature, 
  surface gravity, and reduced $\chi^2$ values are also given in each panel.
  Since no parallax measurements are available for these stars, we
  assume $\logg=8.0$ throughout.\label{fig:fits_photo_exemple}}

\figcaption[f3] {Reduced $\chi^2$ distribution obtained with the
  photometric technique using $ugriz$ photometry for the SDSS sample
  and by assuming $\logg=8.0$. Objects with $\chi_{\rm red}^2>15$ are not
  displayed here.\label{fig:chisq}}

\figcaption[f4] {Transmission curves for the SDSS $ugriz$
  filters as measured by Jim Gunn in 2001 (dashed curves) and
  \citet[][solid curves]{Doi2010}; all transmission curves are
  normalized to unity. Both sets include the
  extinction through an air mass of 1.3. Note that the transmission
  curves from Gunn do not include the complete system
  response from atmosphere to detector.\label{fig:DoivsGunn}}

\figcaption[f5] {Comparison of effective temperatures
  obtained using the transmission curves from Gunn in 2001 and from
  \citet{Doi2010} for the SDSS sample. Objects for which
  $\chi_{\rm red}^2>\chi^2_{\rm crit}$ are shown in red. The dashed line
  represents the 1:1 correspondence.\label{fig:comp_DoivsGunn}}

\figcaption[f6] {Comparison of effective temperatures
  obtained with the photometric technique using $ugriz$ and $griz$
  photometry for the SDSS sample. Surface gravities are assumed to be
  $\logg=8.0$. Objects for which $\chi_{\rm red}^2 > \chi^2_{\rm crit}$ are
  shown in red. The thick solid line represents the 1:1
  correspondence.\label{fig:comp_ugrizvsgriz}}

\figcaption[f7] {Comparison of effective temperatures
  obtained with the photometric technique by assuming $\logg=8.0$ and
  by adopting the surface gravity determined spectroscopically, for the
  SDSS sample. Objects for which $\chi_{\rm red}^2 > \chi^2_{\rm crit}$ are
  shown in red. The dashed line represents the 1:1
  correspondence.\label{fig:comp_gspectro8}}

\figcaption[f8] {Histograms showing the difference between the
  observed (obs) magnitudes and those predicted by the photometric
  technique (th). SDSS to AB$_{95}$ corrections from
  \citet{eisenstein06} have been applied. Only objects with
  $\Te<20,000~{\rm K}$ are considered here. The thick dashed lines correspond to
  $m_{\nu,{\rm obs}}=m_{\nu,{\rm th}}$.\label{fig:histo}}

\figcaption[f9] {Same as Figure \ref{fig:histo} but
  without the SDSS to AB$_{95}$ corrections applied.\label{fig:histo_corrections}}

\figcaption[f10] {Differences between the observed (obs)
  magnitudes and those predicted by the photometric technique (th) as
  a function of the observed magnitude. SDSS to AB$_{95}$ corrections from
  \citet{eisenstein06} have been applied. The red circles represent
  objects with $\Te>20,000~{\rm K}$. The dashed lines correspond to
  $m_{\nu,{\rm obs}}=m_{\nu,{\rm th}}$.\label{fig:dispersion}}

\figcaption[f11] {Distribution of white dwarf distances
  in the SDSS derived from the photometric technique but by adopting
  the spectroscopic $\logg$ values rather than assuming $\logg=8$. Objects for which
  $\chi_{\rm red}^2>\chi^2_{\rm crit}$ are not considered
  here.\label{fig:distances_sdss}}

\figcaption[f12] {Comparison between effective
  temperatures obtained with undereddened and dereddened magnitudes,
  for the SDSS sample. The thick solid line represents the 1:1
  correspondence. Objects for which $\chi_{\rm red}^2 > \chi^2_{\rm crit}$ are
  shown in red.\label{fig:red_vs_dered}}

\figcaption[f13] {Examples of fits obtained with the spectroscopic
  technique. The black lines show the observed spectrum, while the red
  lines correspond to the model fit. Lines range from H$\beta$
  (bottom) to H8 (top) and are offset vertically for clarity. The
  effective temperature and surface gravity (uncorrected for 3D
  effects) of each star are also given. Examples of problematic SDSS
  spectra are shown at the bottom.\label{fig:ex_spectro}}

\figcaption[f14] {Top panel: Atmospheric parameters
  for the SDSS spectroscopic sample determined with the spectroscopic
  technique using 1D/MLT models, for $30,000~{\rm K} \ge \Te \ge
  6000~{\rm K}$. Bottom panel: Same results but with the 3D hydrodynamical corrections
  from \citet{TremblayIV}.
  The dashed line in both panels shows a 0.592 \msun\ evolutionary
  track, which corresponds to the median mass (corrected) for this sample in this
  temperature range. \label{fig:corr_spectro_SDSS}}

\figcaption[f15] {Same as Figure
  \ref{fig:corr_spectro_SDSS} but for the Gianninas sample. The
  dashed line shows a 0.616 \msun\ evolutionary track, which
  corresponds to the median mass (corrected) for this sample in this temperature
  range.\label{fig:corr_spectro_Gianninas}}

\figcaption[f16] {Differences in surface gravity
  (top panel) and effective temperature (bottom panel) between 1D/MLT
  models (${\rm ML}2/\alpha=0.7$) and 3D hydrodynamical models from
  \citet{TremblayIV}, for $7.0< \logg < 9.0$ and $20,000~{\rm K} > \Te
  > 5000~{\rm K}$.\label{fig:functions_Tremblay}}

\figcaption[f17] {Mass distributions for the SDSS spectroscopic sample
  obtained without (left) and with (right) the 3D correction functions
  from \citet{TremblayIV}. The black histograms show the total sample,
  while the blue ones correspond to objects with $\Te>13,000~{\rm K}$
  and the red ones to objects with $\Te < 13,000~{\rm K}$. The mean
  masses and standard deviations (in solar mass units) of the
  corresponding samples are also given.\label{fig:histo_mass_SDSS}}

\figcaption[f18] {Same as Figure \ref{fig:histo_mass_SDSS} but for the
  Gianninas spectroscopic sample.\label{fig:histo_mass_Gianninas}}

\figcaption[f19] {Comparison of effective temperatures
  obtained with the spectroscopic method between the SDSS and
  Gianninas samples. Correction functions from \citet{TremblayIV} have
  been applied. The dashed line represents the 1:1
  correspondence.\label{fig:comp_spec}}

\figcaption[f20] {Comparison between
  effective temperatures obtained from photometry and spectroscopy for
  the SDSS sample. Magnitudes have been corrected for interstellar
  reddening. Objects for which $\chi_{\rm red}^2>\chi^2_{\rm crit}$ are shown in
  red. The thick solid line represents the 1:1
  correspondence.\label{fig:comp_phot_SDSS_rougissement_g2}}

\figcaption[f21] {Same as Figure
  \ref{fig:comp_phot_SDSS_rougissement_g2} but magnitudes have not
  been corrected for interstellar reddening.\label{fig:comp_spectro_photo_log}}

\figcaption[f22] {Left panel: Comparison
  between effective temperatures obtained from photometry and
  spectroscopy for the Gianninas subset (see text). Magnitudes have
  been corrected for interstellar reddening. Objects for which
  $\chi_{\rm red}^2>\chi^2_{\rm crit}$ are shown in red. The thick solid line
  represents the 1:1 correspondence. Right panel: Same as the left panel
  but for the SDSS sample.\label{fig:comp_G_SDSS}}

\figcaption[f23] {Comparison of effective temperatures obtained from
  photometry and spectroscopy for the Gianninas subset using Stark
  profiles from \citet{Lemke1997} and twice the value of the critical
  microfield $\beta_{\rm crit}$ (see text). Magnitudes have been
  corrected for interstellar reddening. Objects for which
  $\chi_{\rm red}^2>\chi^2_{\rm crit}$ are shown in red. The thick solid line
  represents the 1:1
  correspondence.\label{fig:spec_photo_old_rouge_g}}

\figcaption[f24] {Atmospheric parameters for the SDSS
  sample obtained from photometry (black circles) by exploiting the
  sensitivity of the Balmer jump ($u-g$ color index) to $\logg$
  between 8000 K and 17,000 K. Magnitudes have been corrected for
  interstellar reddening. Objects for which $\chi_{\rm red}^2>\chi^2_{\rm crit}$ are not
  shown here. The dashed line shows a 0.592 \msun\ evolutionary track,
  which is the median mass for the SDSS sample in this temperature
  range. The red dots show the spectroscopic results for the same
  objects.\label{fig:logg_phot_dered}}

\figcaption[f25] {Photometric (black) and spectroscopic (red) $\logg$
  distributions for white dwarfs in the SDSS with $8000~{\rm
    K}<\Te<17,000~{\rm K}$. The spectroscopic $\logg$ values have been
  corrected for 3D hydrodynamical
  effects.\label{fig:histo_logg_dered}}

\clearpage

\begin{figure}[p]
\plotone{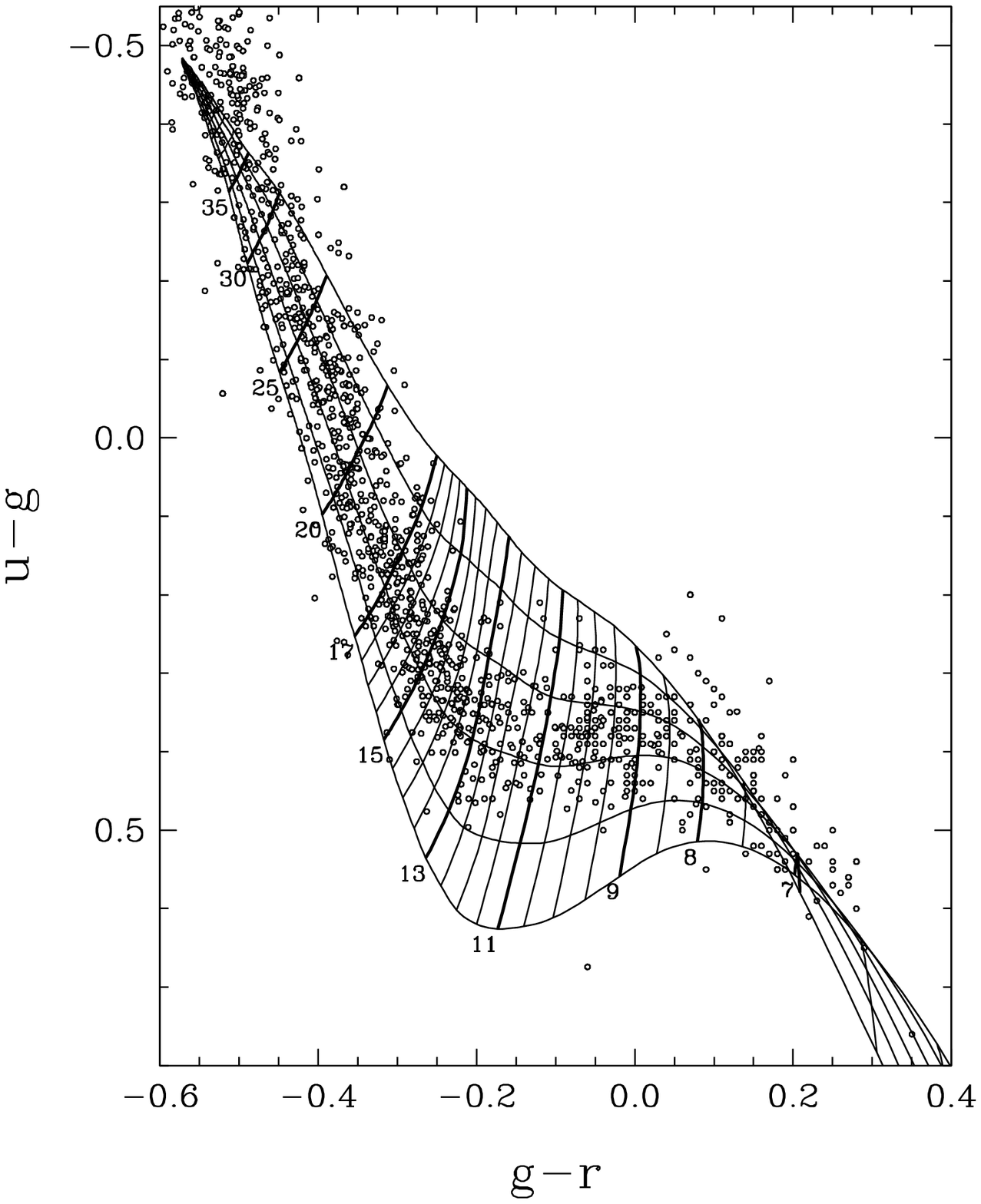}
\begin{flushright}
Figure \ref{fig:umg_gmr}
\end{flushright}
\end{figure}

\clearpage

\begin{figure}[p]
\plotone{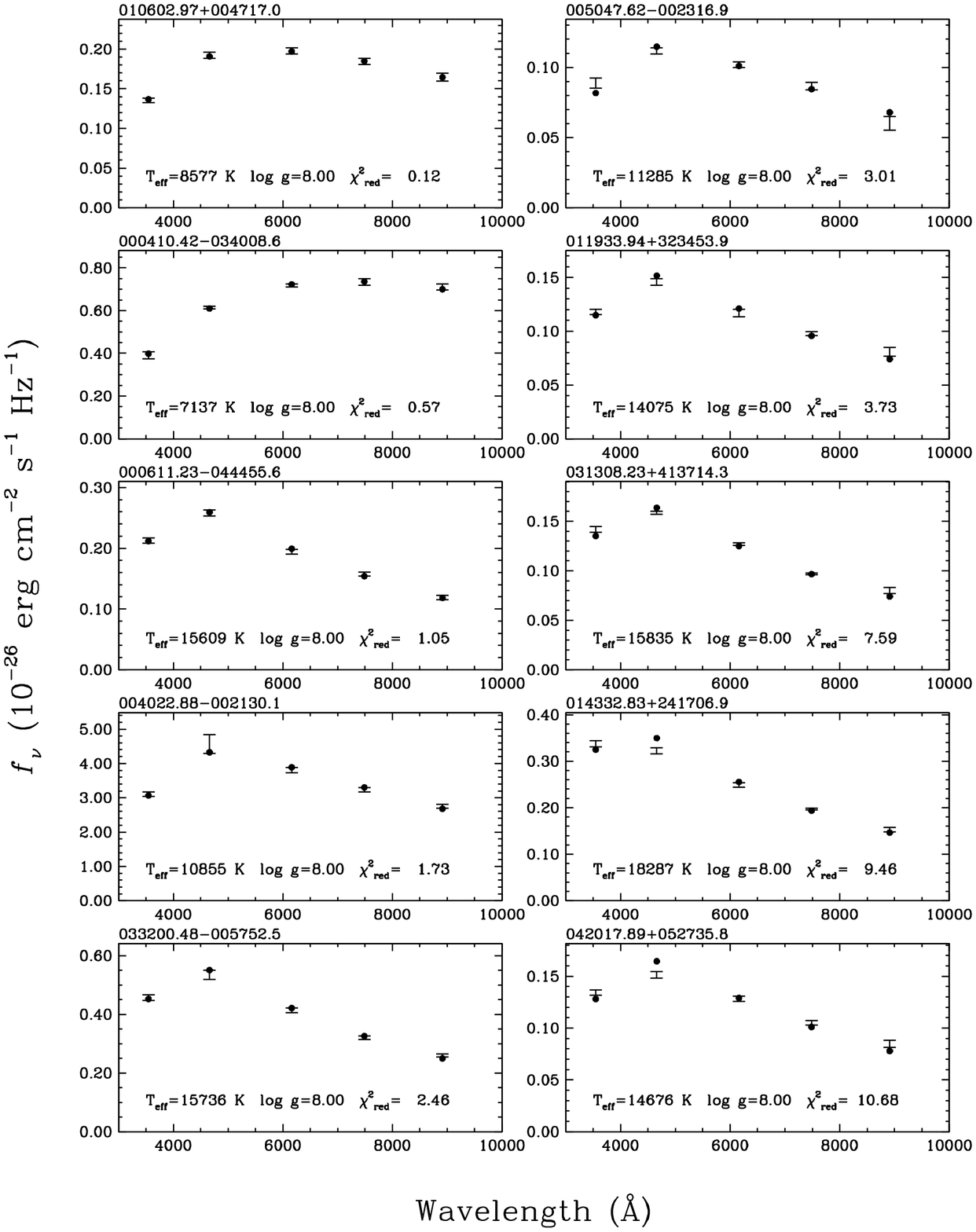}
\begin{flushright}
Figure \ref{fig:fits_photo_exemple}
\end{flushright}
\end{figure}

\clearpage

\begin{figure}[p]
\plotone{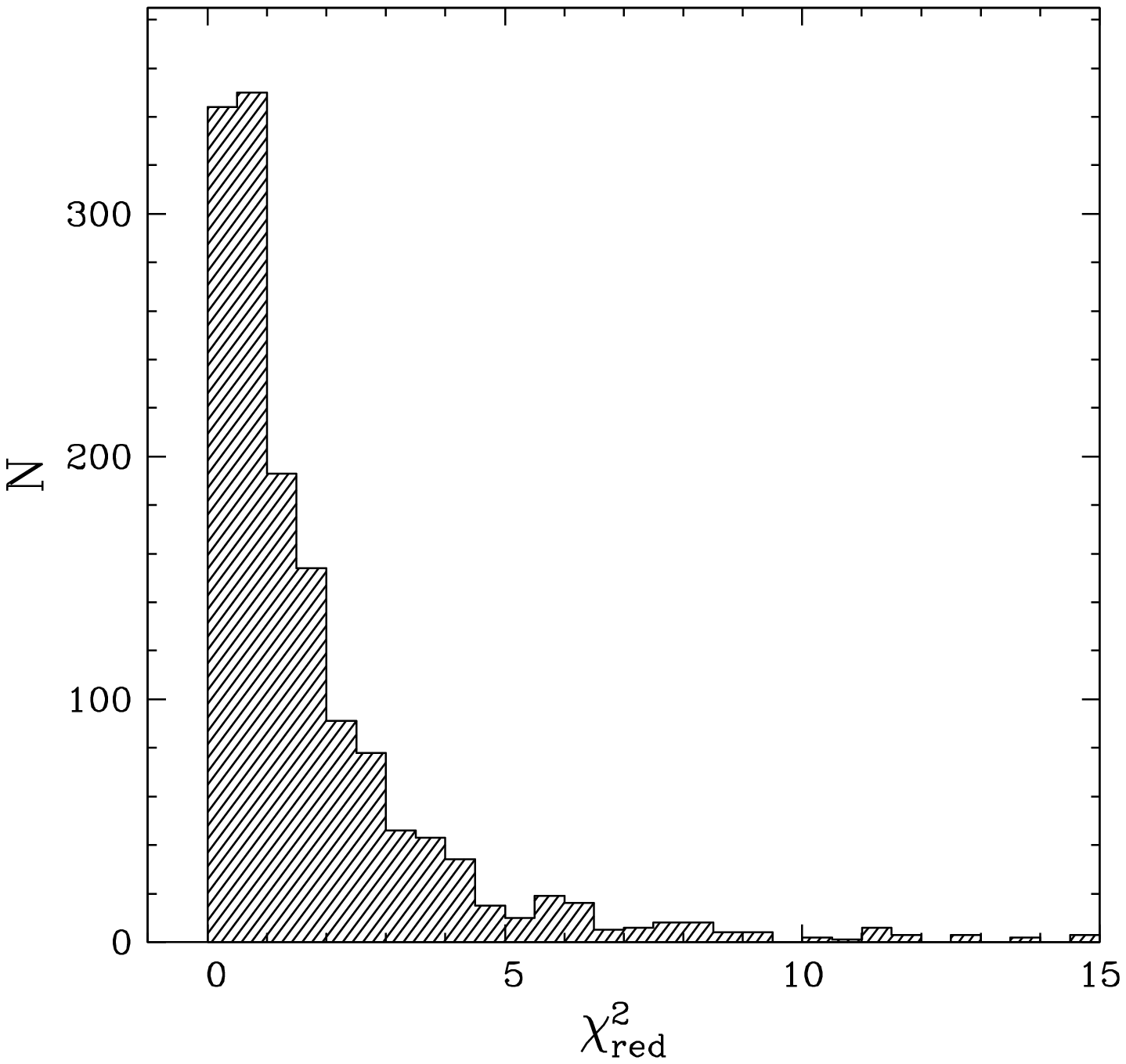}
\begin{flushright}
Figure \ref{fig:chisq}
\end{flushright}
\end{figure}

\clearpage

\begin{figure}[p]
\plotone{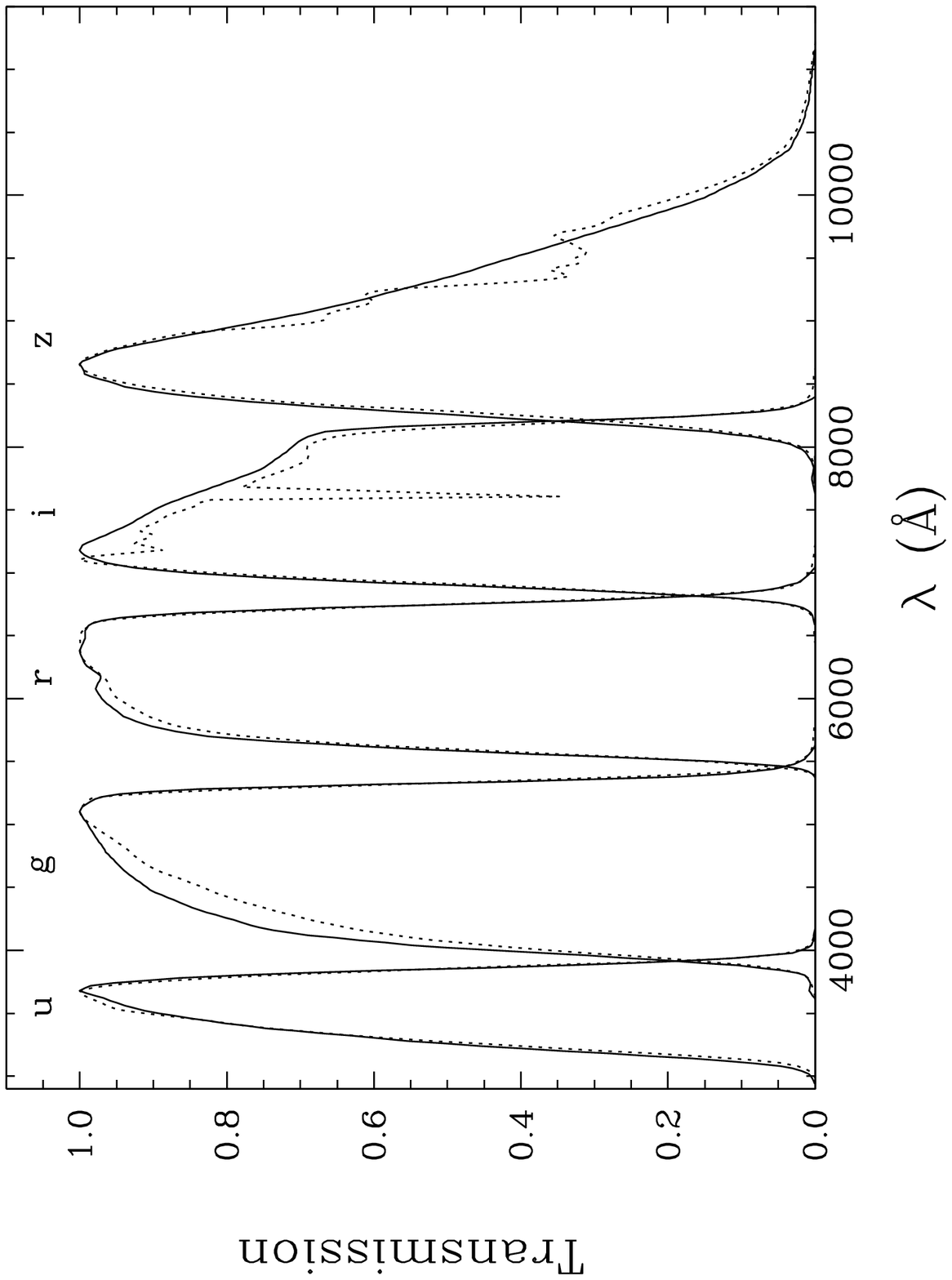}
\begin{flushright}
Figure \ref{fig:DoivsGunn}
\end{flushright}
\end{figure}

\clearpage

\begin{figure}[p]
\plotone{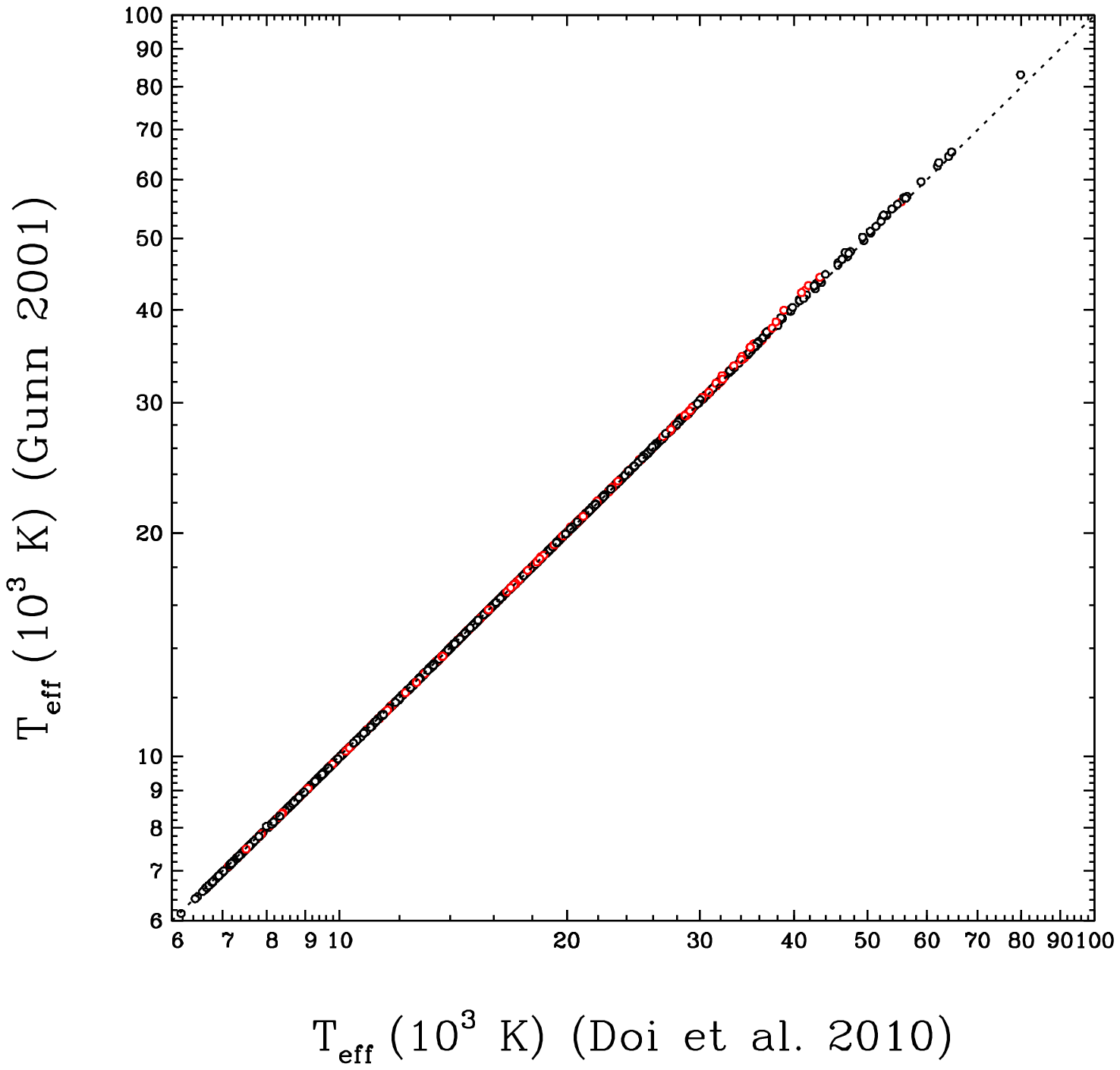}
\begin{flushright}
Figure \ref{fig:comp_DoivsGunn}
\end{flushright}
\end{figure}

\clearpage

\begin{figure}[p]
\plotone{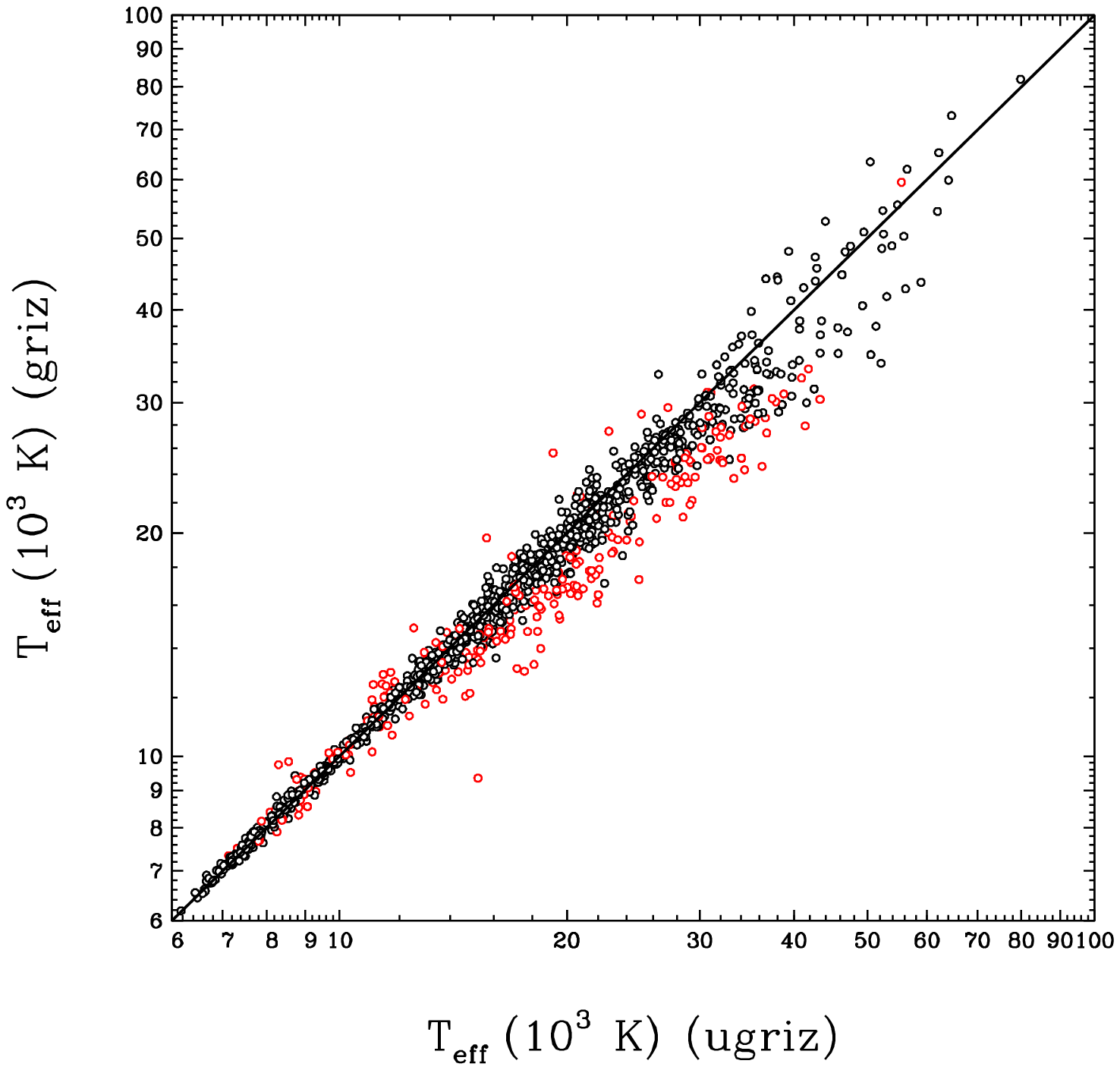}
\begin{flushright}
Figure \ref{fig:comp_ugrizvsgriz}
\end{flushright}
\end{figure}

\clearpage

\begin{figure}[p]
\plotone{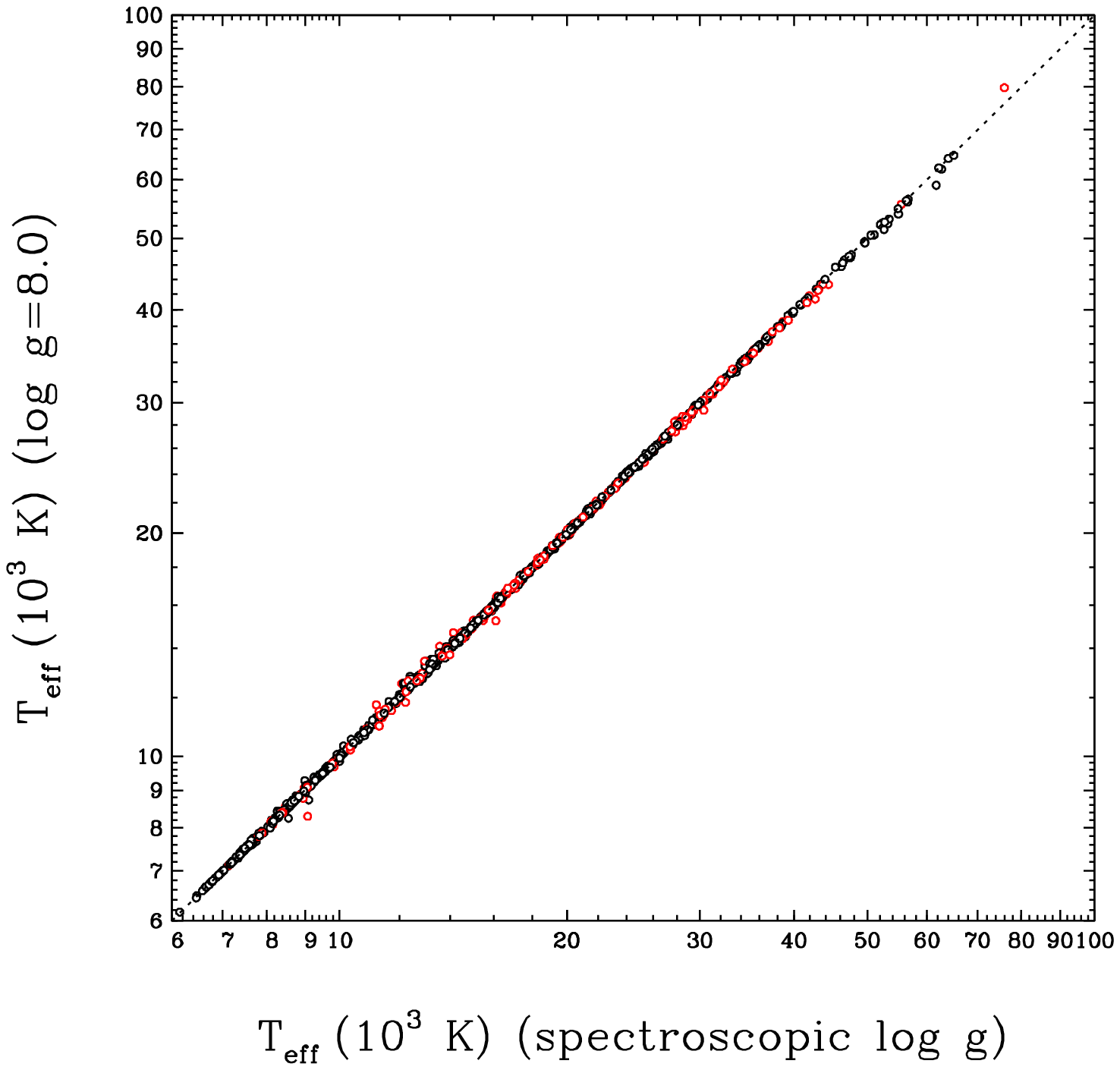}
\begin{flushright}
Figure \ref{fig:comp_gspectro8}
\end{flushright}
\end{figure}

\clearpage

\begin{figure}[p]
\plotone{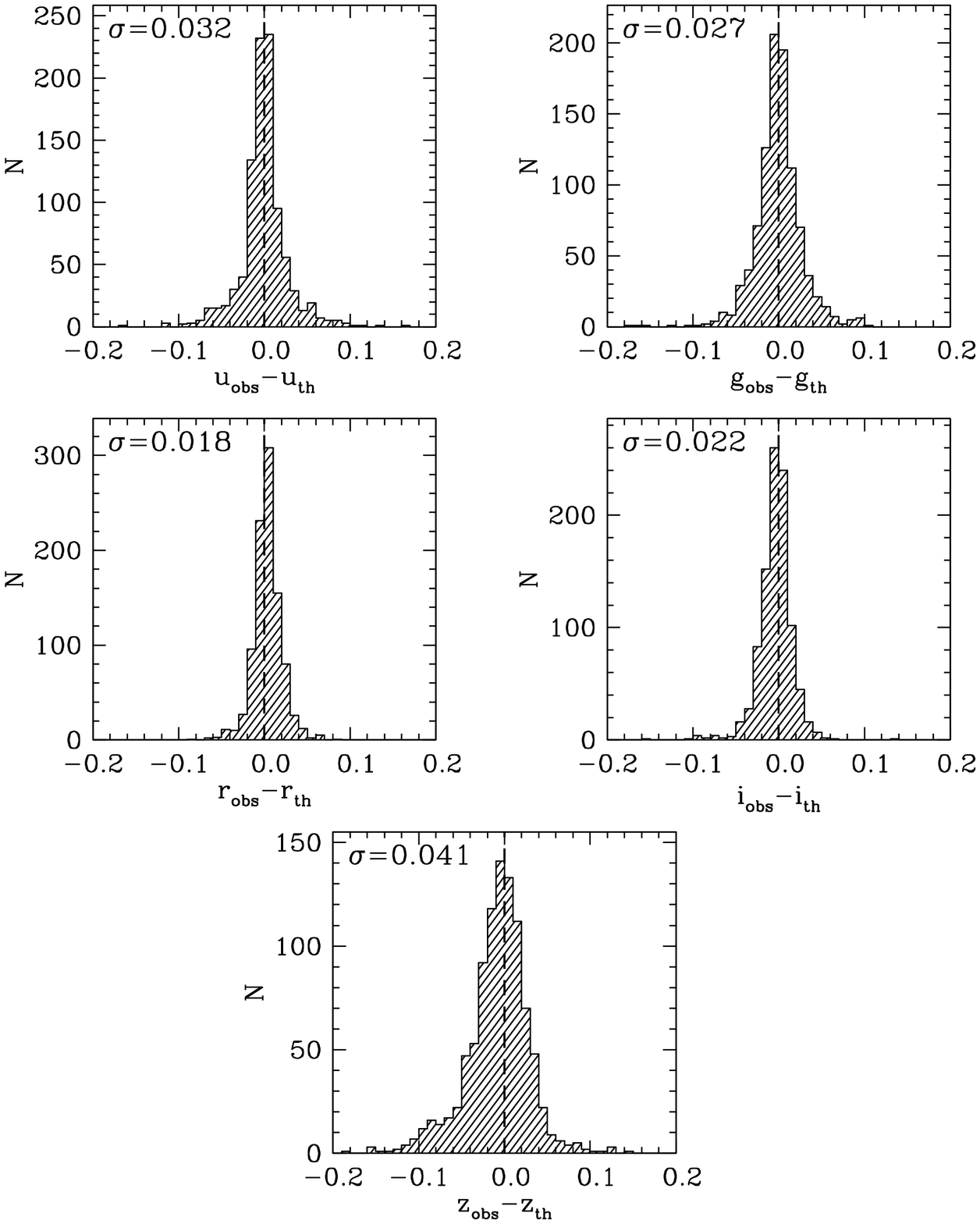}
\begin{flushright}
Figure \ref{fig:histo}
\end{flushright}
\end{figure}

\clearpage
\begin{figure}[p]
\plotone{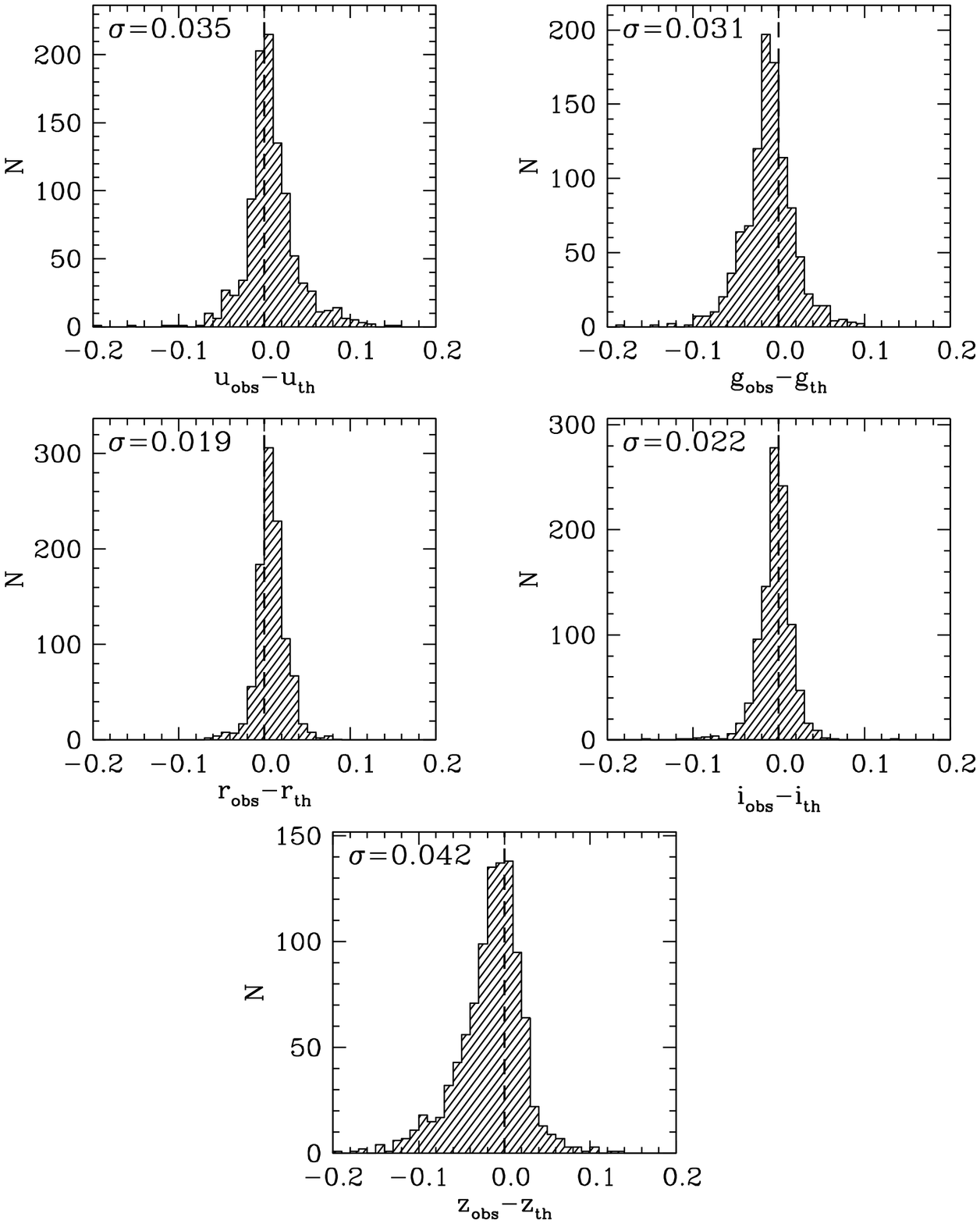}
\begin{flushright}
Figure \ref{fig:histo_corrections}
\end{flushright}
\end{figure}

\clearpage

\begin{figure}[p]
\plotone{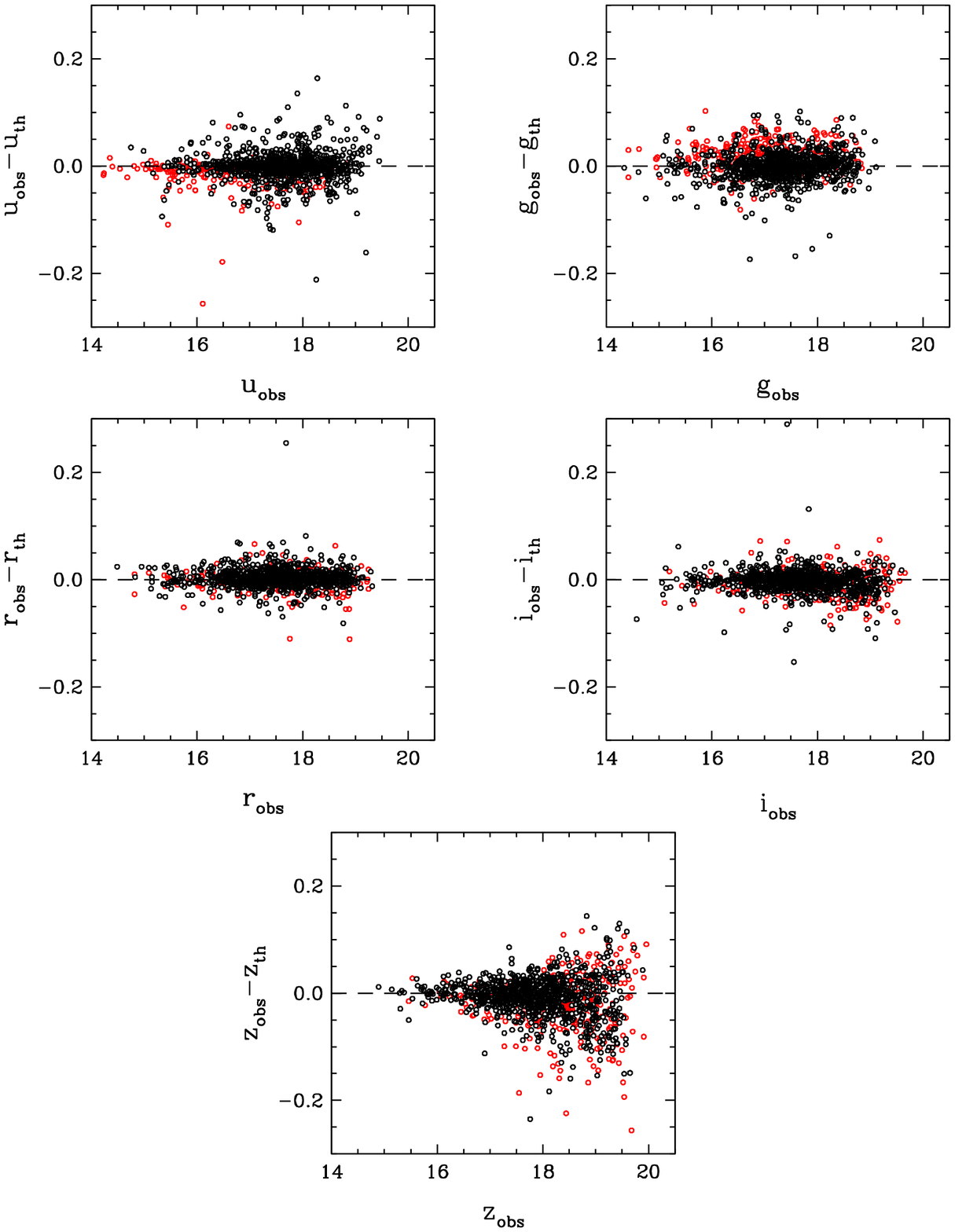}
\begin{flushright}
Figure \ref{fig:dispersion}
\end{flushright}
\end{figure}

\clearpage

\begin{figure}[p]
\plotone{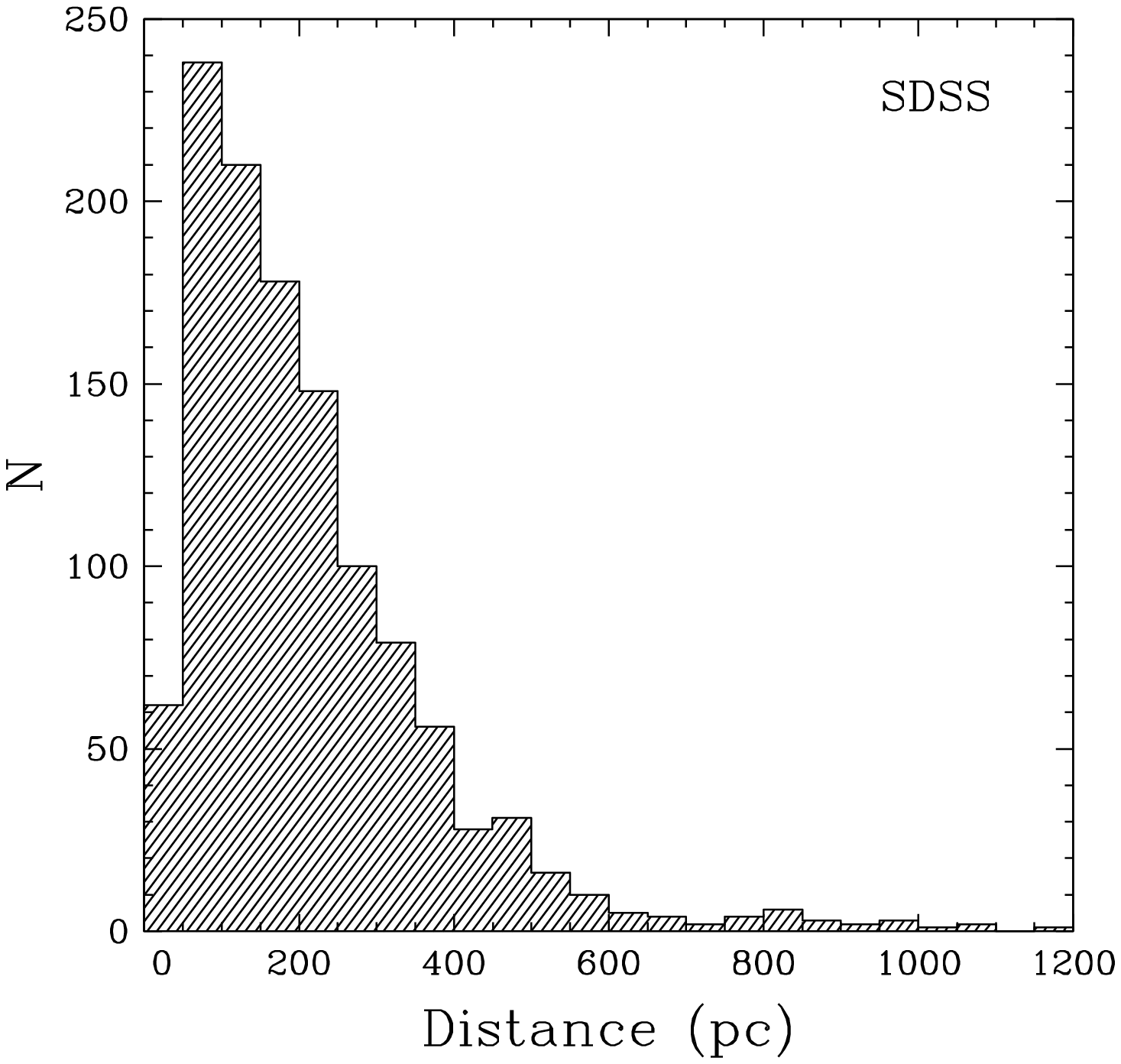}
\begin{flushright}
Figure \ref{fig:distances_sdss}
\end{flushright}
\end{figure}

\clearpage

\begin{figure}[p]
\plotone{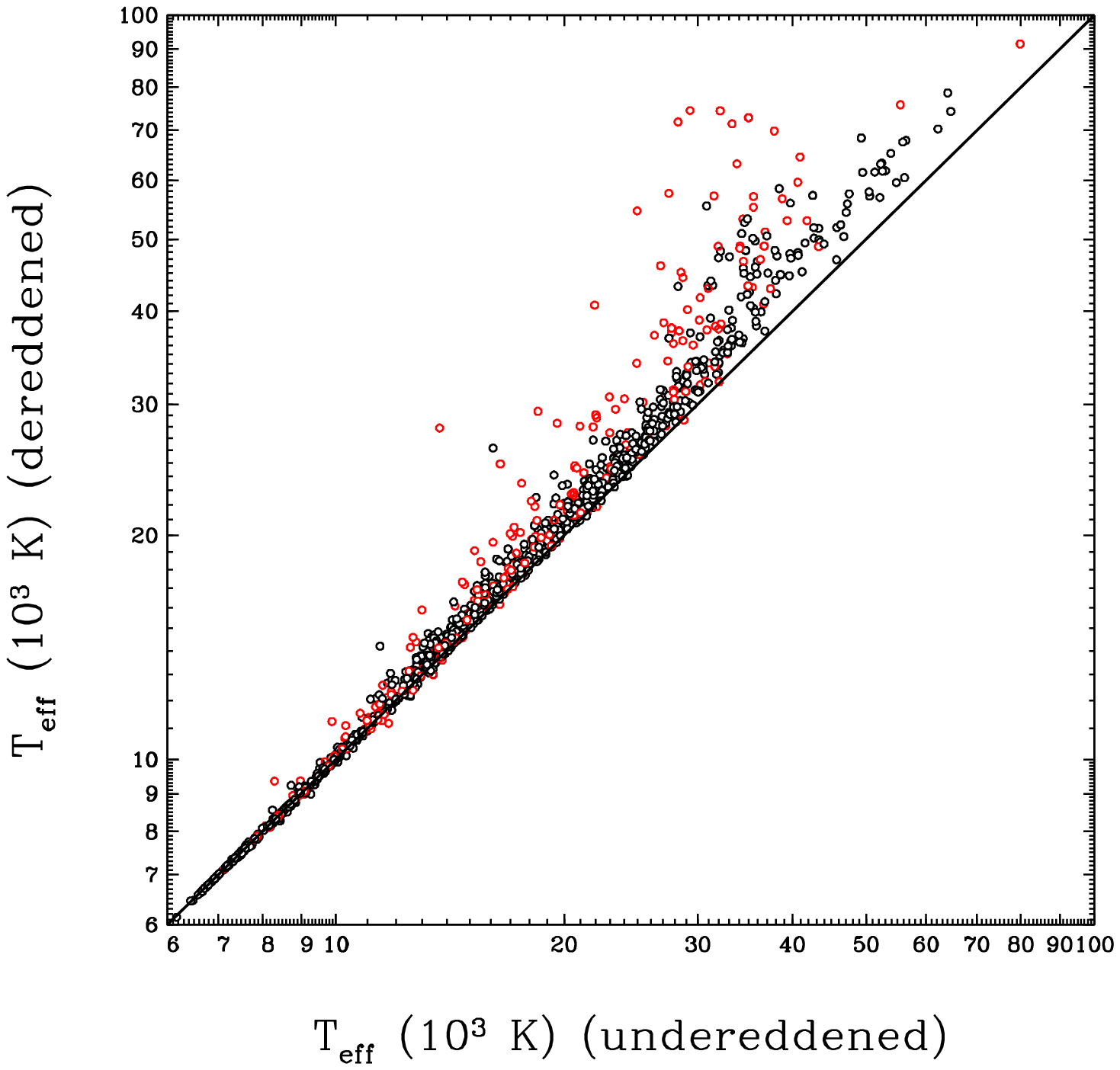}
\begin{flushright}
Figure \ref{fig:red_vs_dered}
\end{flushright}
\end{figure}

\clearpage
\begin{figure}[p]
\plotone{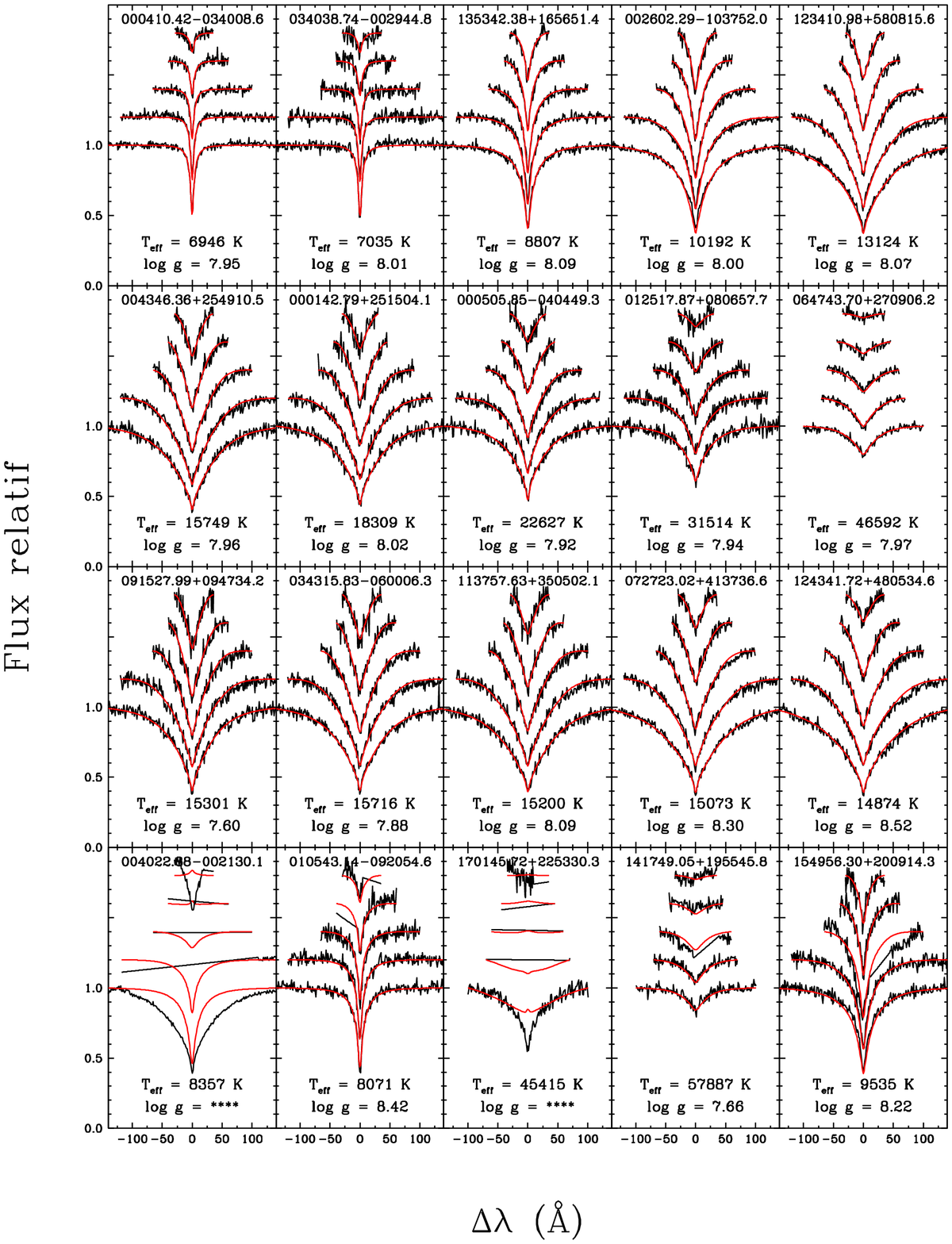}
\begin{flushright}
Figure \ref{fig:ex_spectro}
\end{flushright}
\end{figure}

\clearpage

\begin{figure}[p]
\plotone{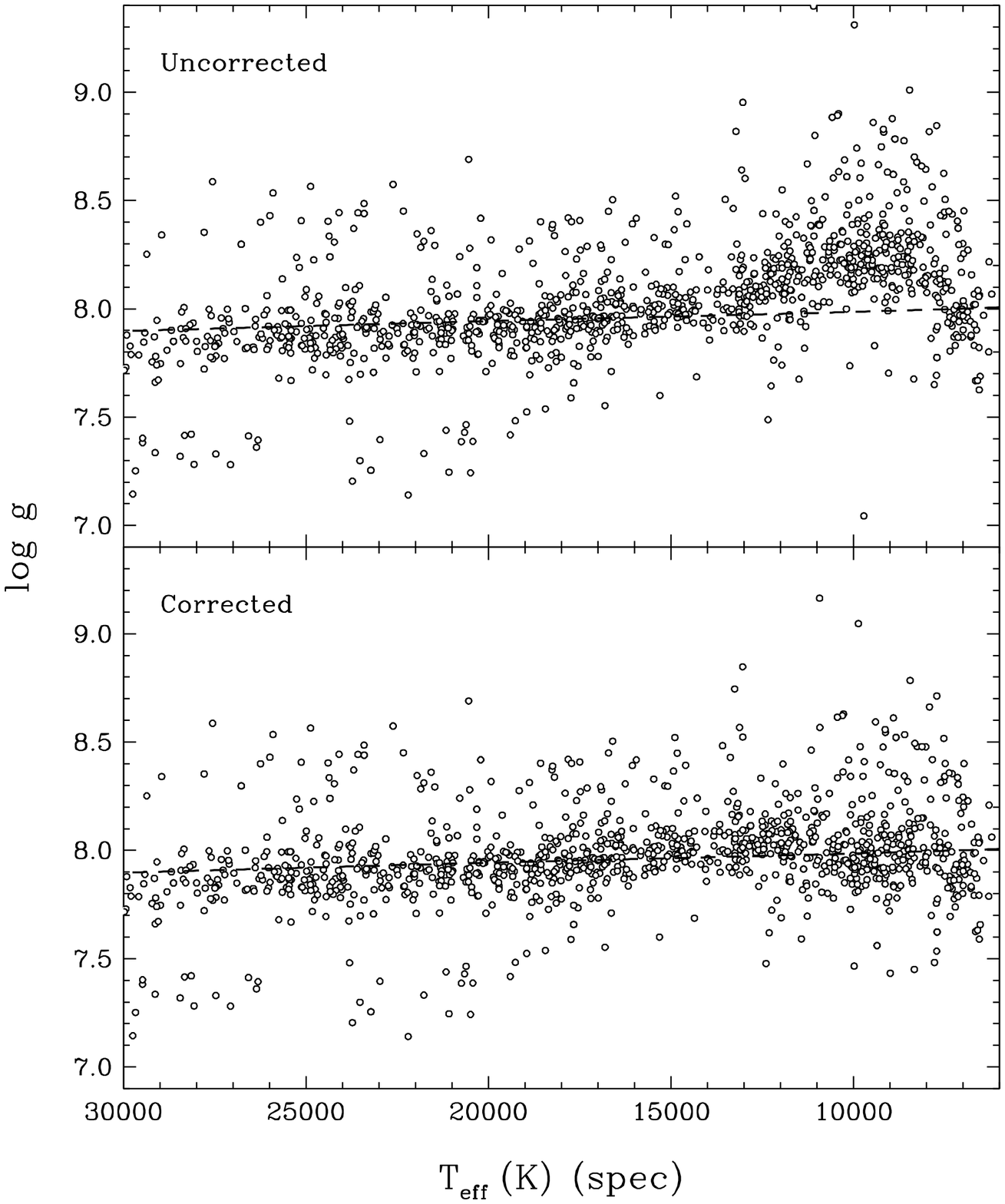}
\begin{flushright}
Figure \ref{fig:corr_spectro_SDSS}
\end{flushright}
\end{figure}

\clearpage

\begin{figure}[p]
\plotone{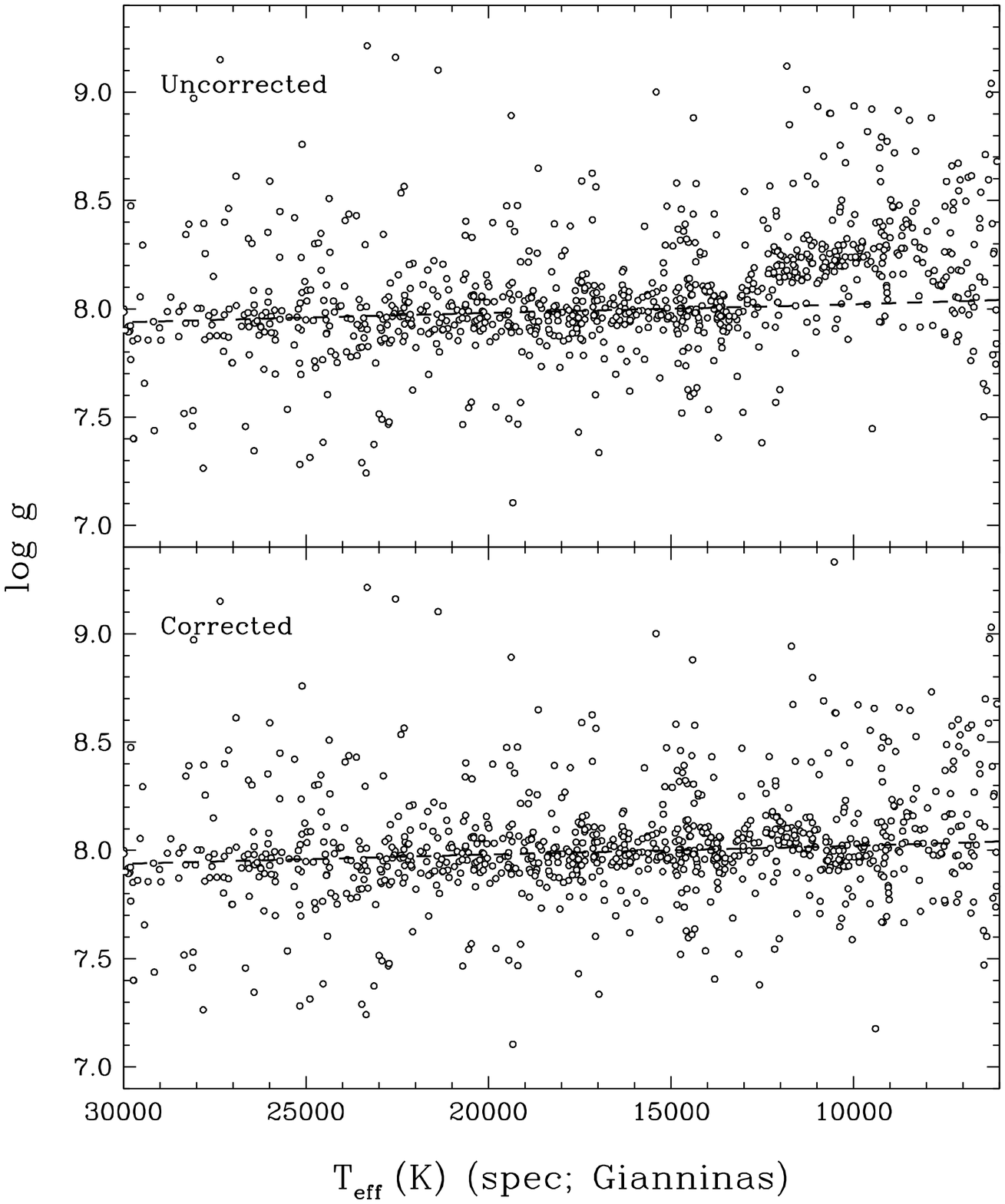}
\begin{flushright}
Figure \ref{fig:corr_spectro_Gianninas}
\end{flushright}
\end{figure}

\clearpage

\begin{figure}[p]
\plotone{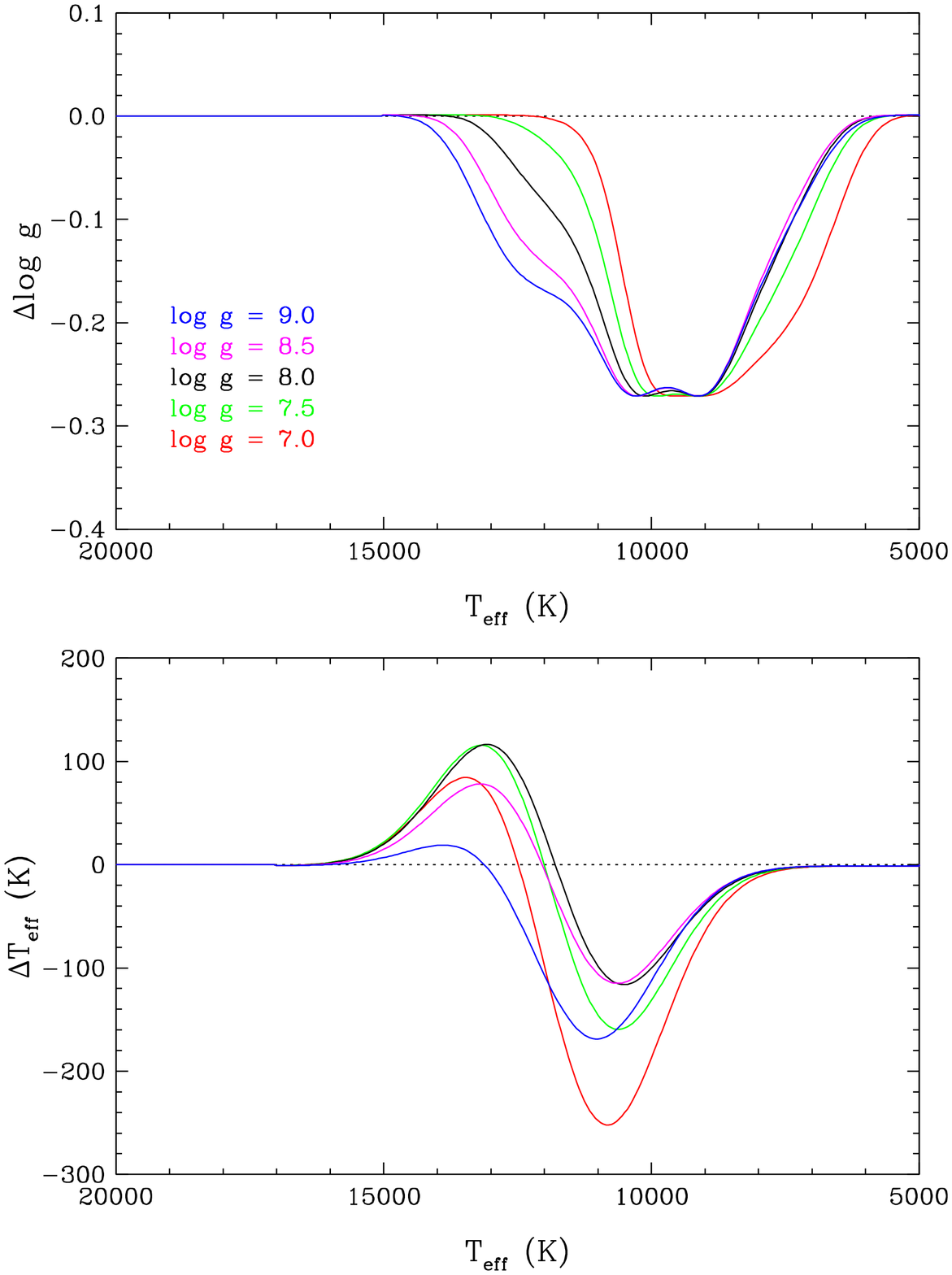}
\begin{flushright}
Figure \ref{fig:functions_Tremblay}
\end{flushright}
\end{figure}

\clearpage

\begin{figure}[p]
\plotone{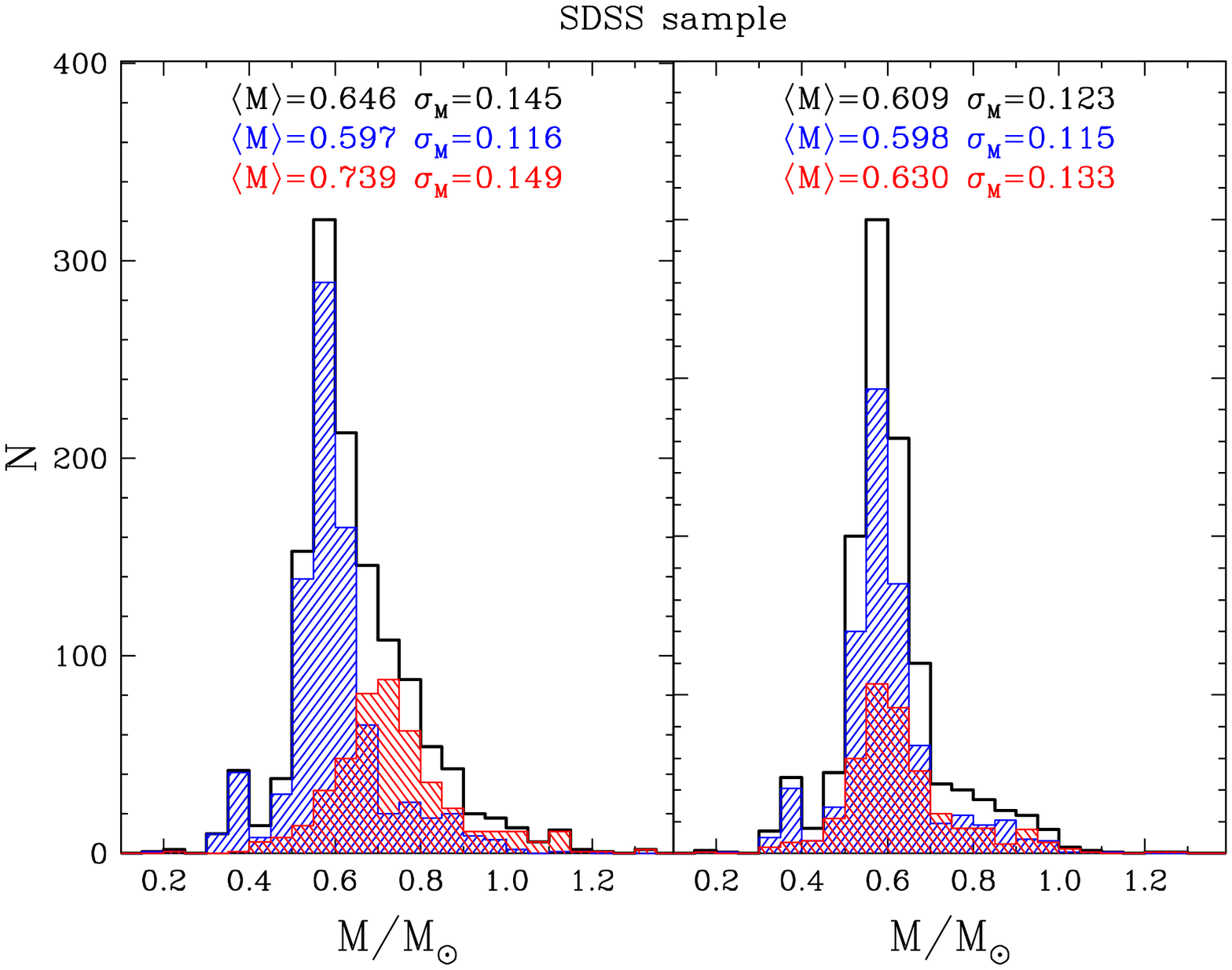}
\begin{flushright}
Figure \ref{fig:histo_mass_SDSS}
\end{flushright}
\end{figure}

\clearpage

\begin{figure}[p]
\plotone{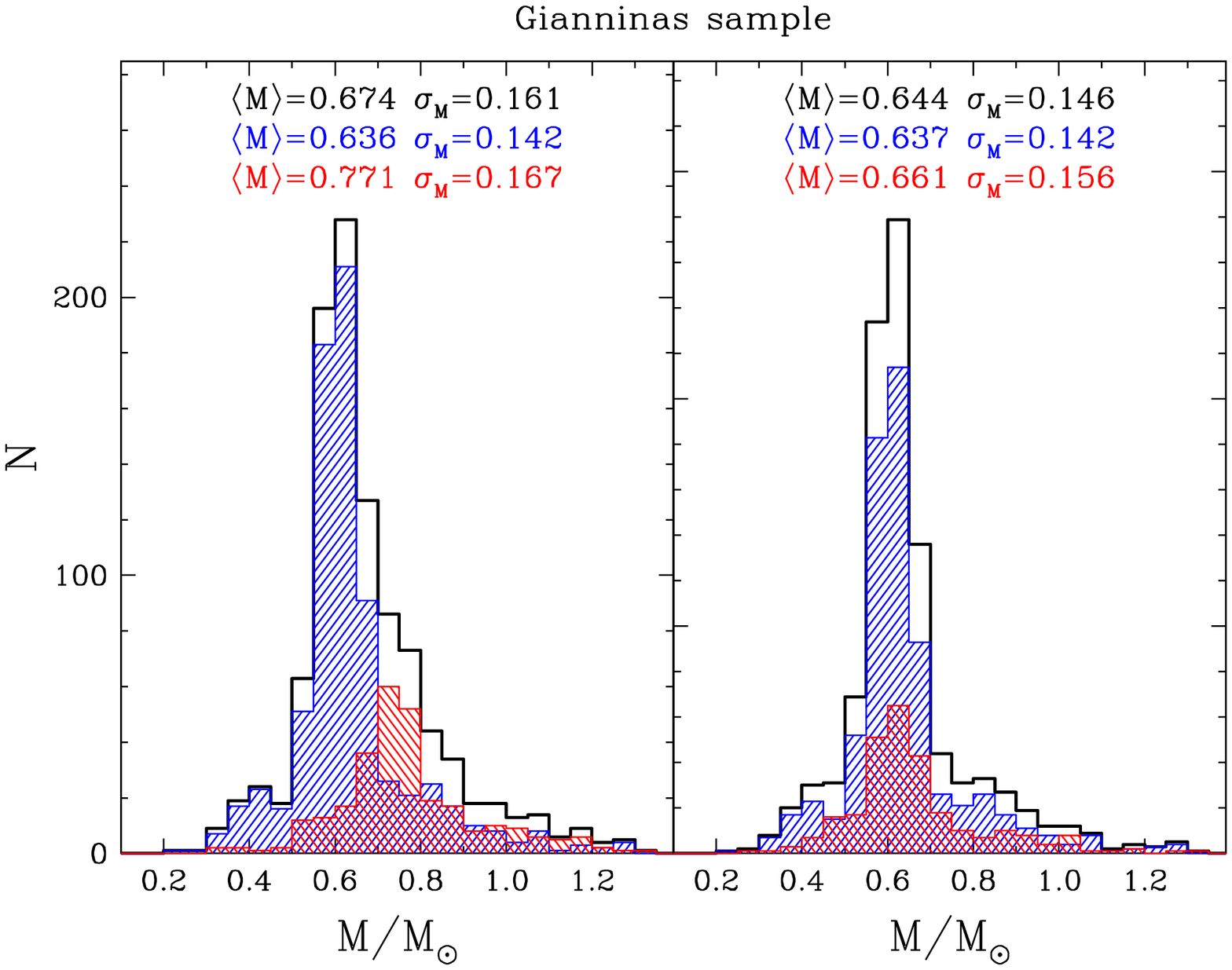}
\begin{flushright}
Figure \ref{fig:histo_mass_Gianninas}
\end{flushright}
\end{figure}

\clearpage

\begin{figure}[p]
\plotone{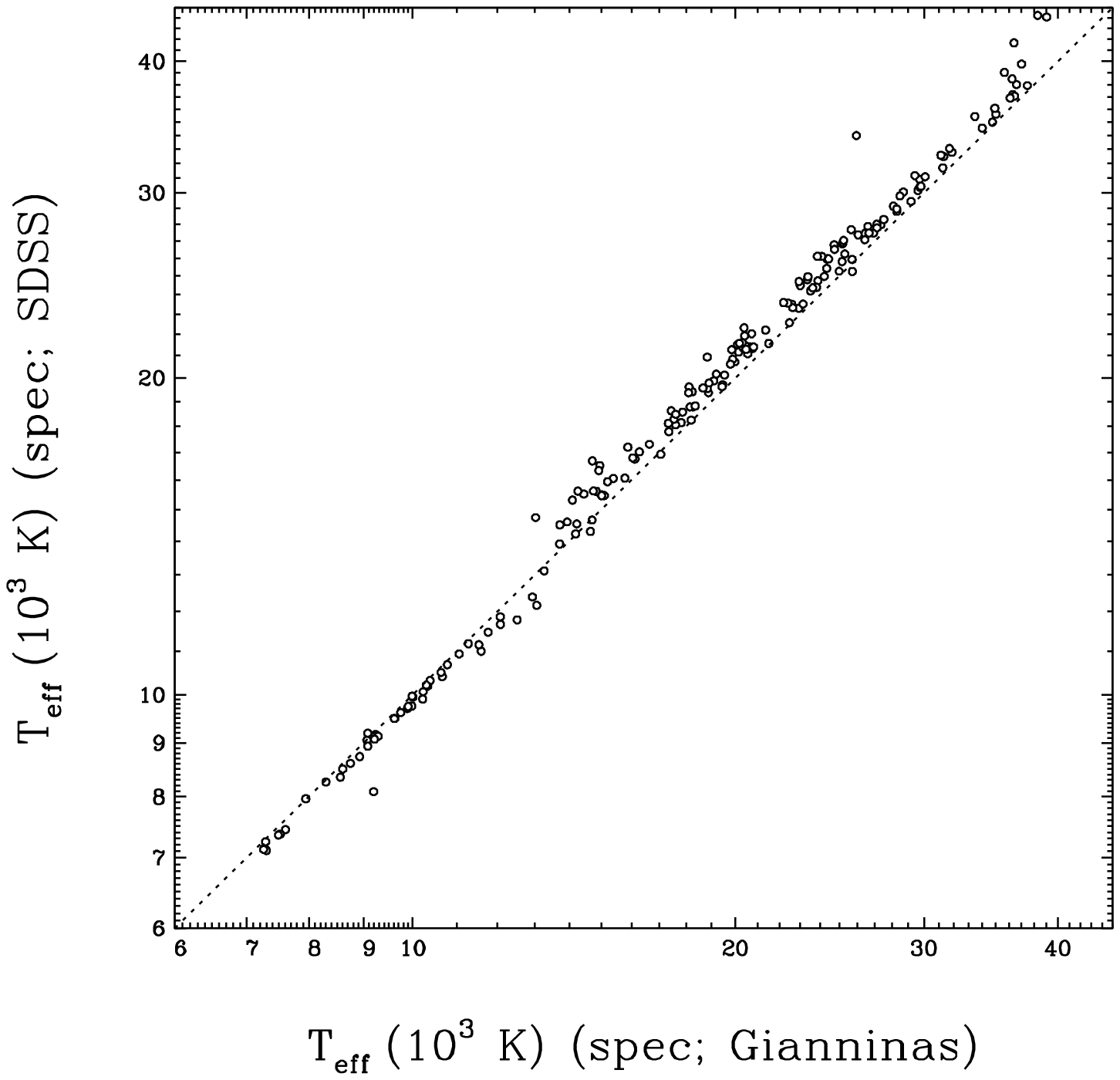}
\begin{flushright}
Figure \ref{fig:comp_spec}
\end{flushright}
\end{figure}

\clearpage

\begin{figure}[p]
\plotone{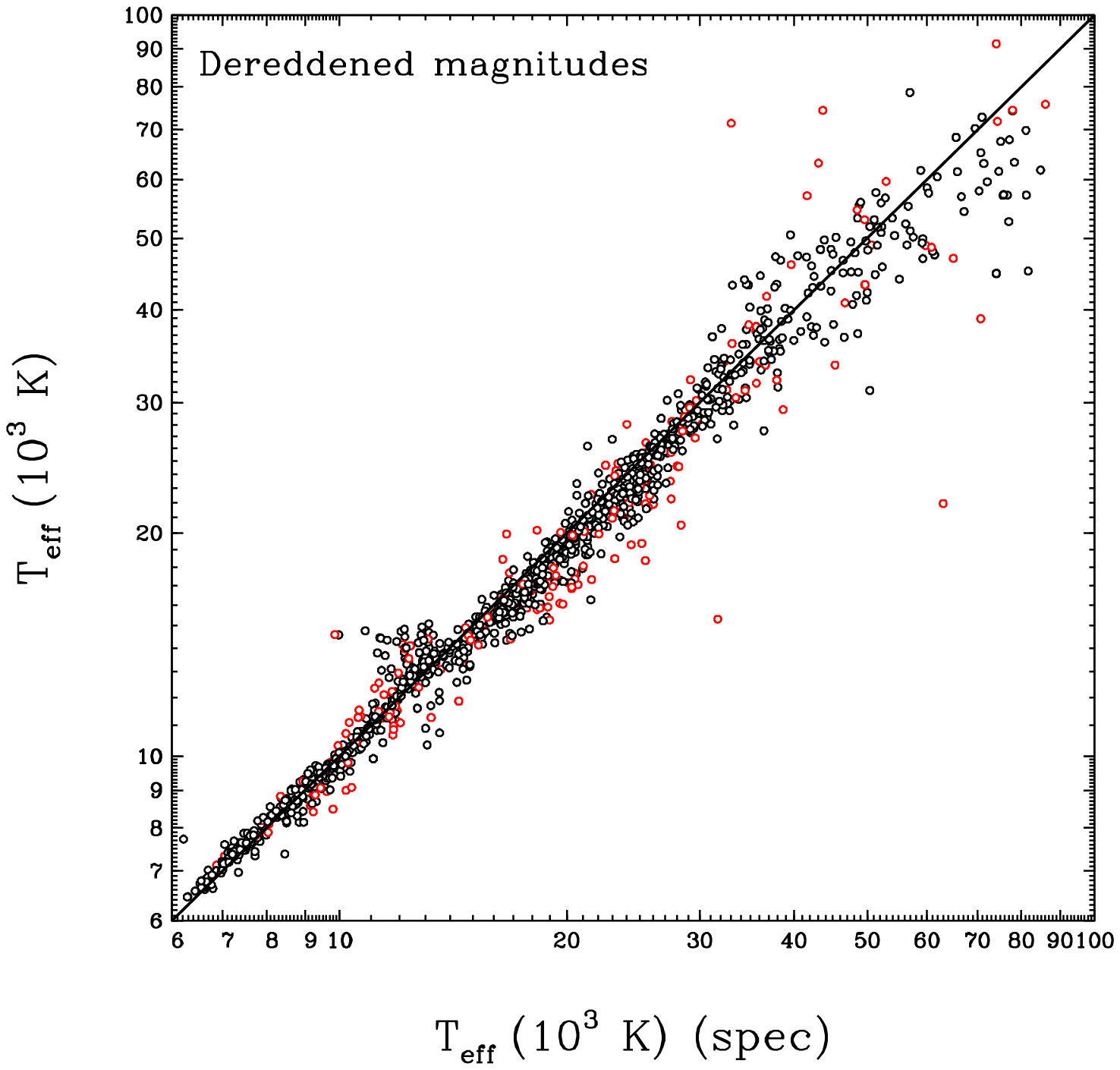}
\begin{flushright}
Figure \ref{fig:comp_phot_SDSS_rougissement_g2}
\end{flushright}
\end{figure}

\clearpage

\begin{figure}[p]
\plotone{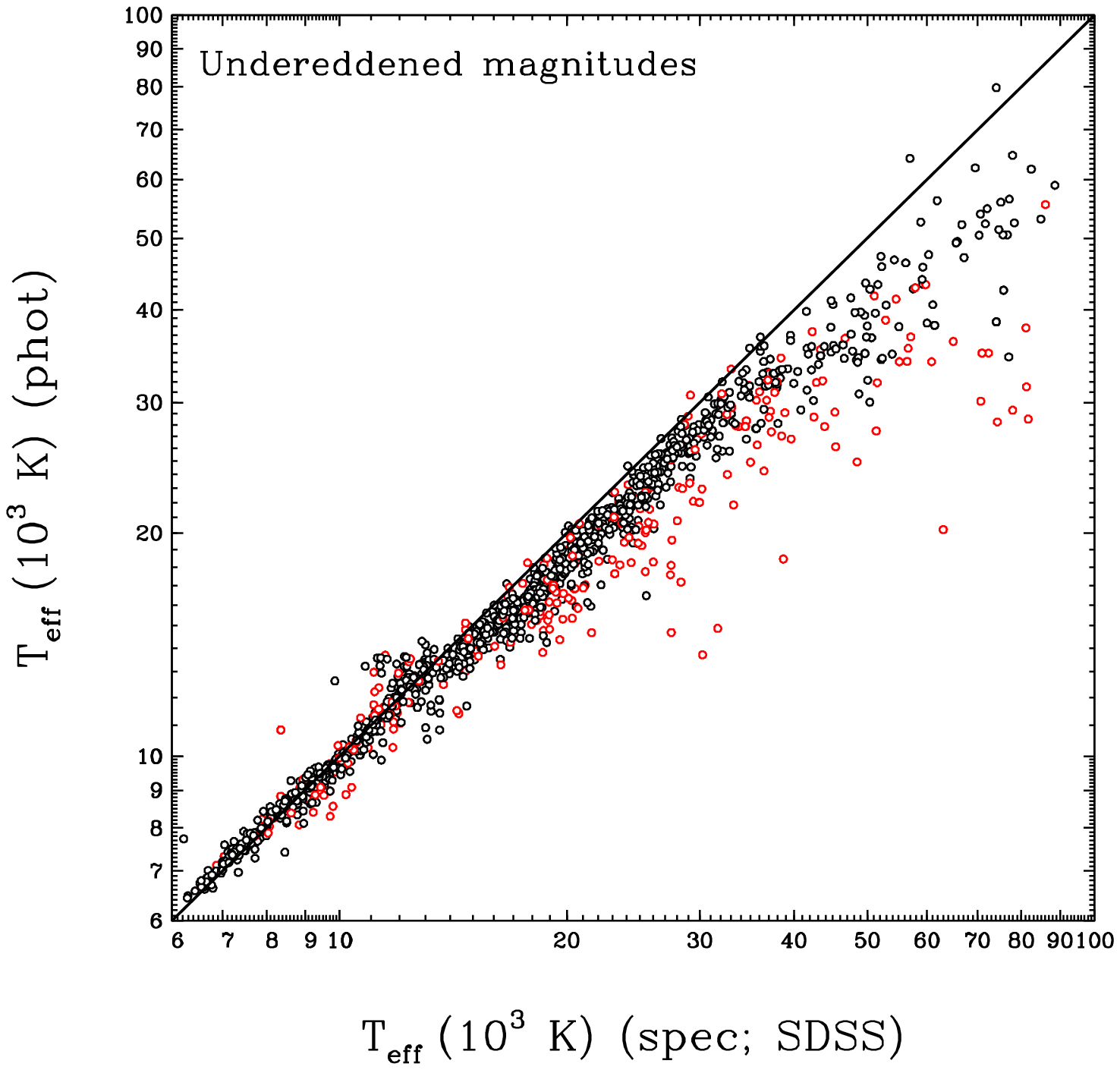}
\begin{flushright}
Figure \ref{fig:comp_spectro_photo_log}
\end{flushright}
\end{figure}

\clearpage

\begin{figure}[p]
\plotone{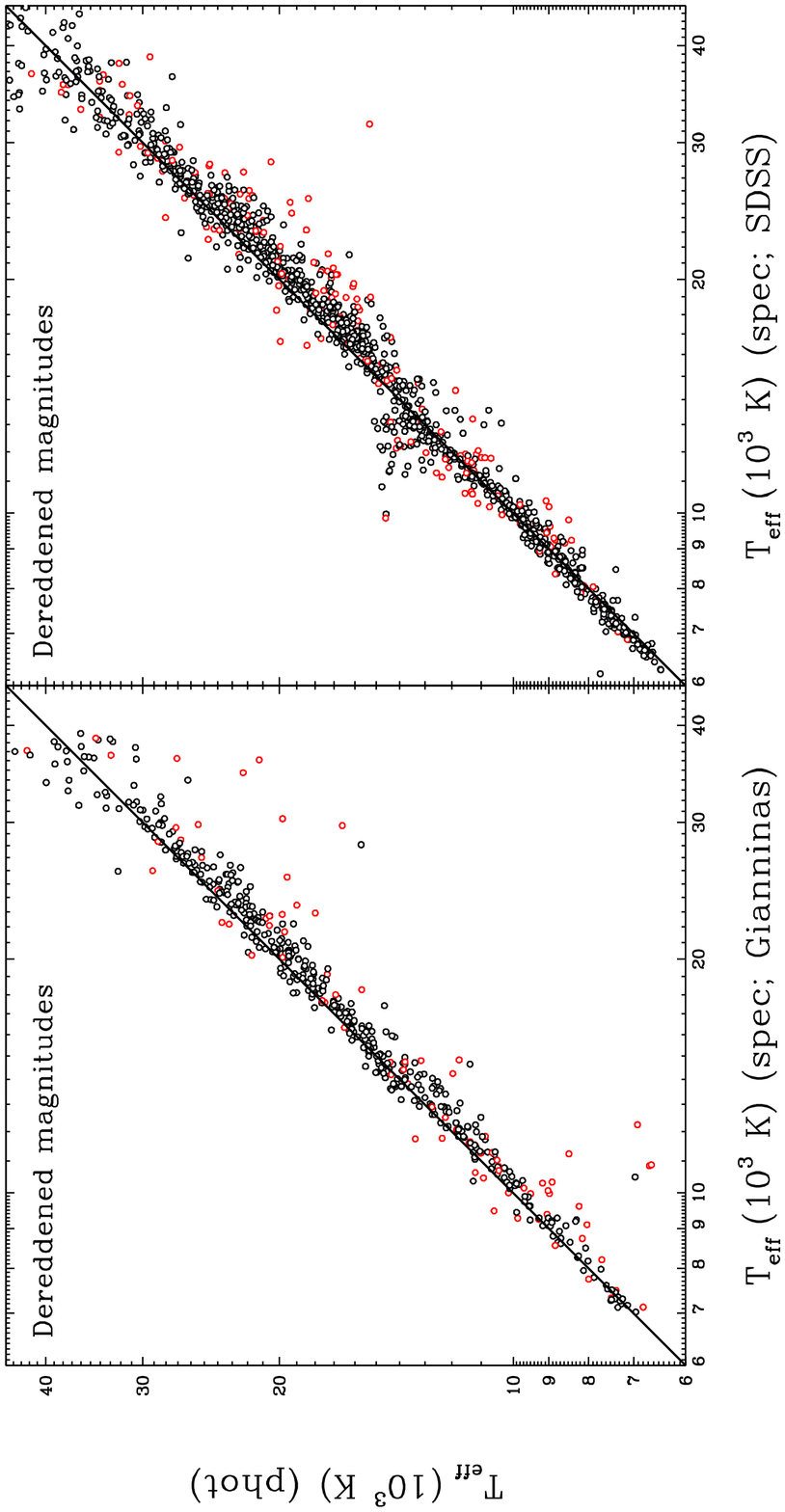}
\begin{flushright}
Figure \ref{fig:comp_G_SDSS}
\end{flushright}
\end{figure}

\clearpage

\begin{figure}[p]
\plotone{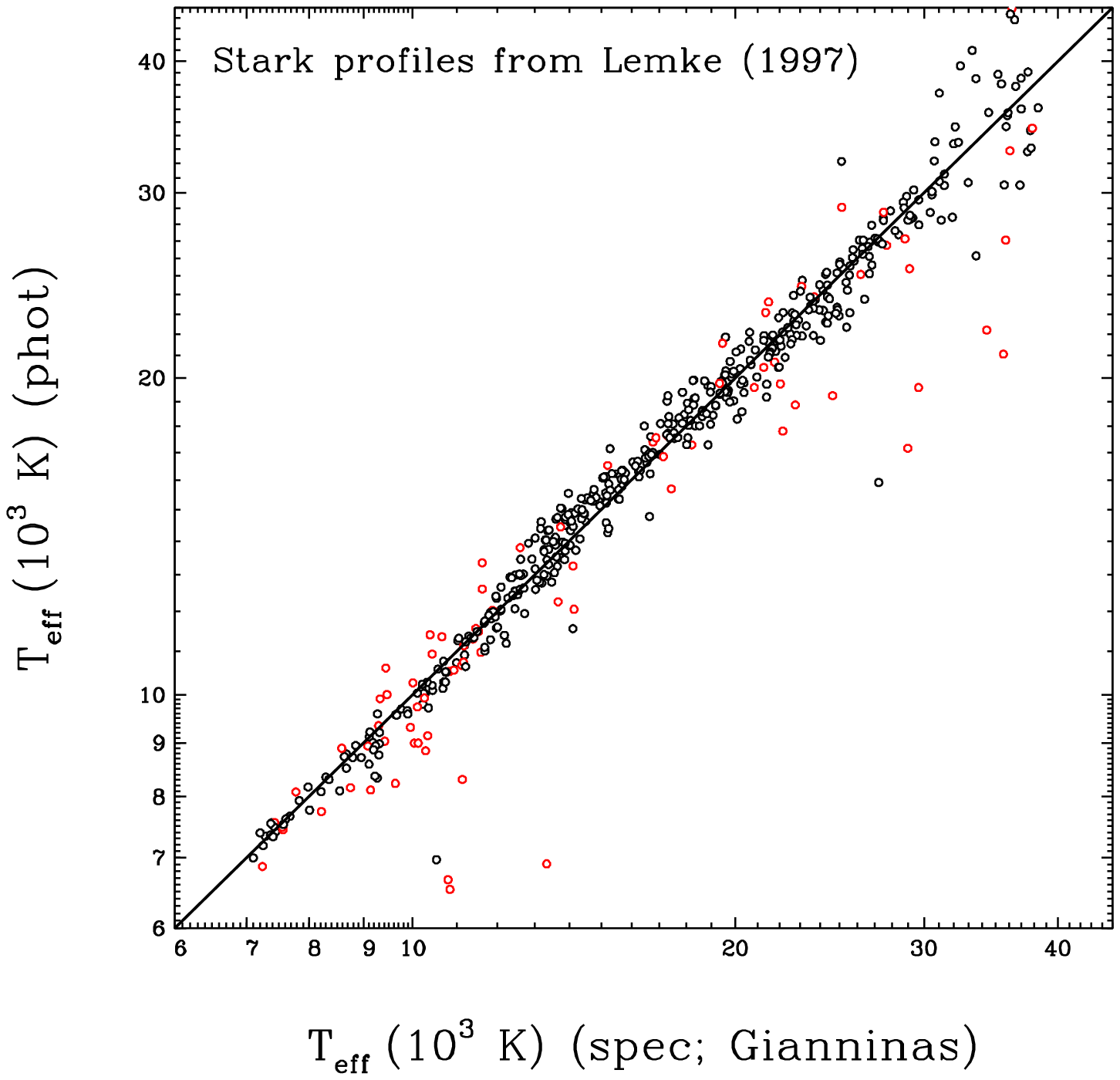}
\begin{flushright}
Figure \ref{fig:spec_photo_old_rouge_g}
\end{flushright}
\end{figure}

\clearpage

\begin{figure}[p]
\plotone{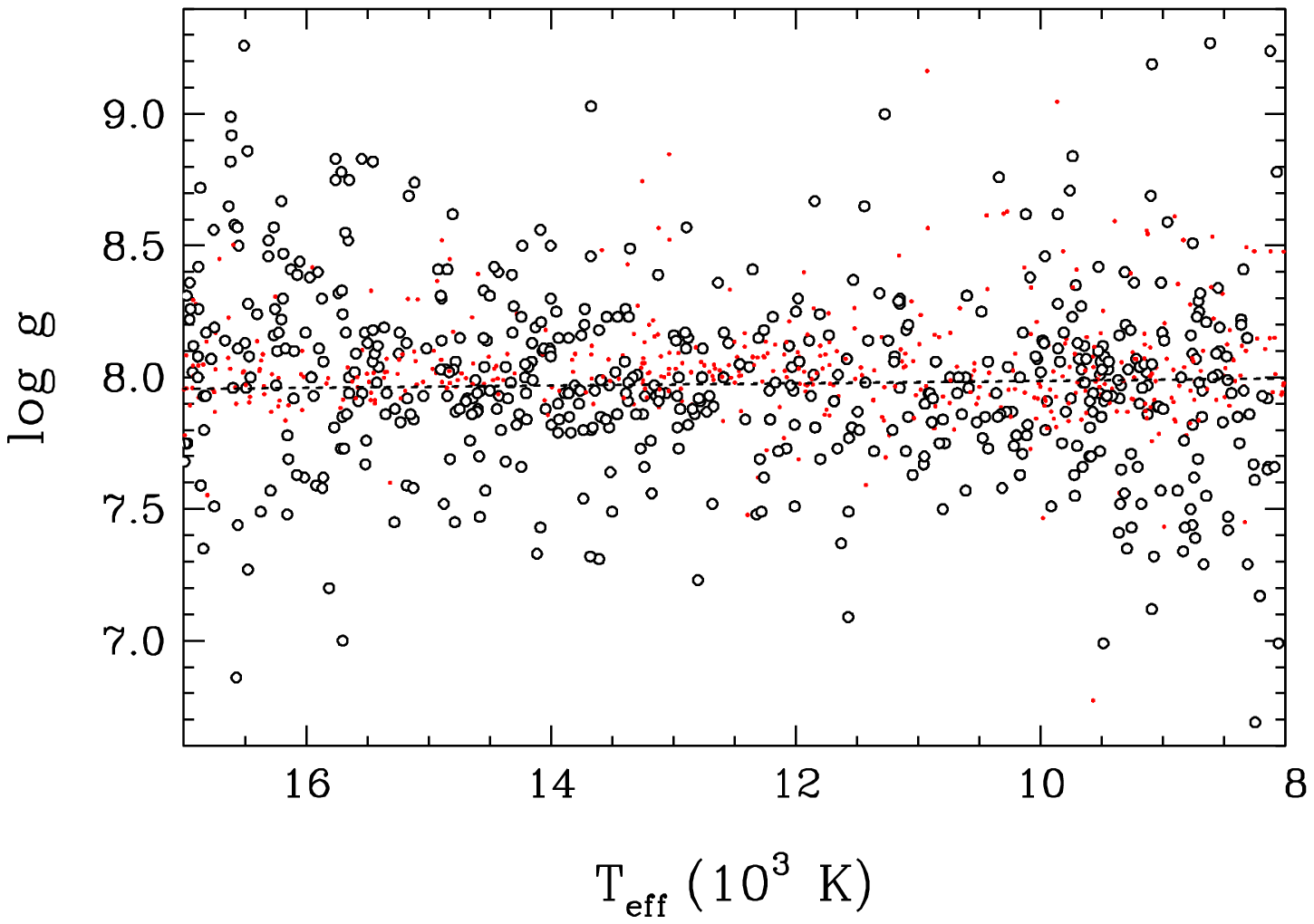}
\begin{flushright}
Figure \ref{fig:logg_phot_dered}
\end{flushright}
\end{figure}

\clearpage

\begin{figure}[p]
\plotone{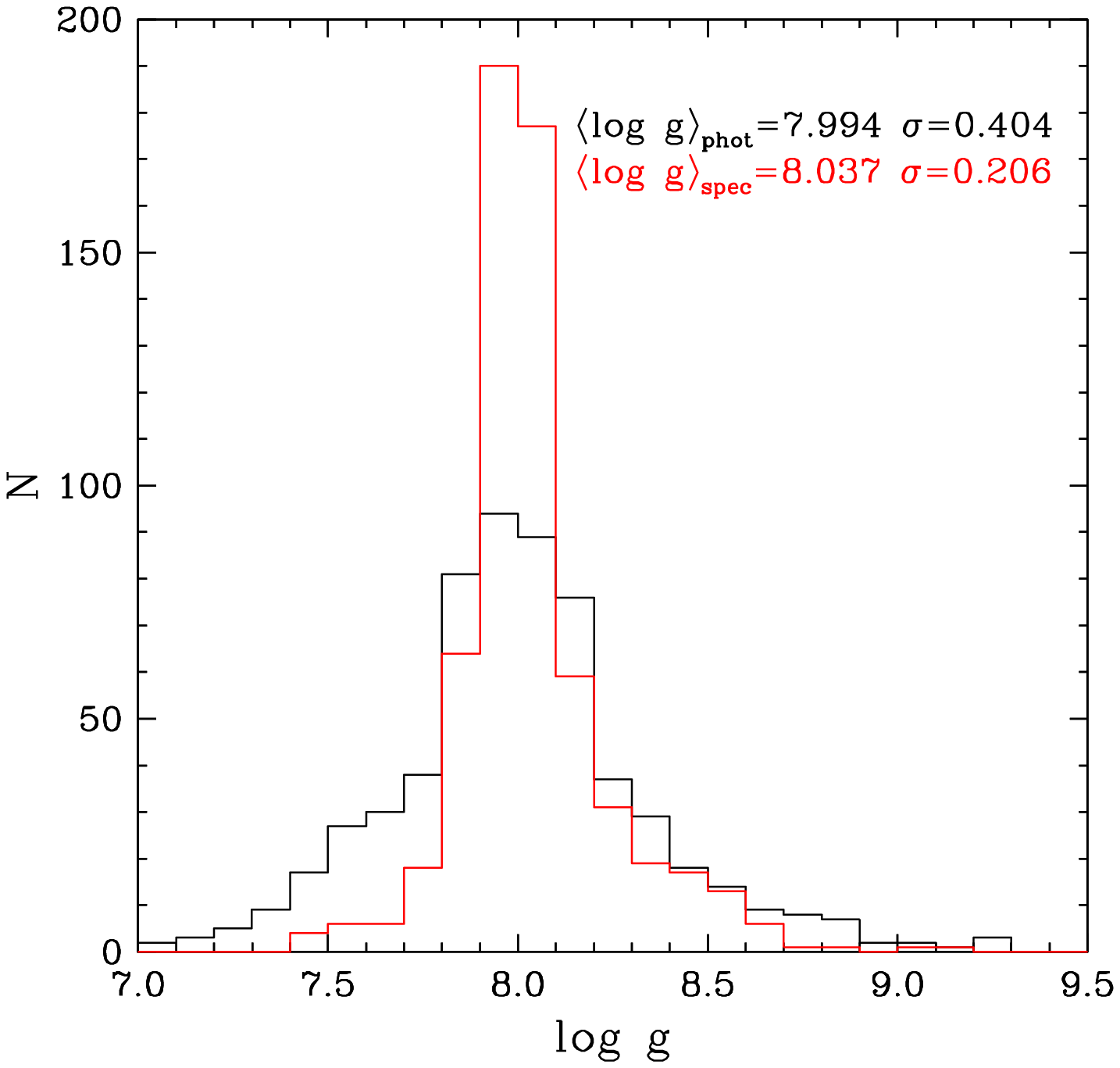}
\begin{flushright}
Figure \ref{fig:histo_logg_dered}
\end{flushright}
\end{figure}

\end{document}